\documentclass[10pt,twocolumn,english,prd,superscriptaddress,nofootinbib,preprintnumbers,showpacs,floatfix]{revtex4-2}

\usepackage{dcolumn}
\usepackage{bm}
\usepackage{longtable}
\usepackage{natbib}
\usepackage{textcase}
\usepackage{url}
\usepackage{relsize}
\usepackage{amsfonts}
\usepackage{amsmath}
\usepackage{amssymb,epsf}
\usepackage{latexsym}
\usepackage{graphicx,epsfig}
\usepackage{amssymb}
\usepackage{float}
\usepackage{subfigure}
\usepackage{epstopdf}
\usepackage[colorlinks=true,citecolor=blue,linkcolor=blue,urlcolor=black]{hyperref}
\usepackage{psfrag}
\usepackage{wrapfig}
\usepackage{makeidx}
\usepackage{epsf}
\usepackage{xcolor}
\usepackage{mathtools}

\begin{document}


\title{Novel scalar degrees of freedom emerging from hybrid metric-Palatini gravity}

\author{David S. Pereira}%
	\email{djpereira@fc.ul.pt}
	\affiliation{%
		Departamento de F\'{i}sica, Faculdade de Ci\^{e}ncias da Universidade de Lisboa, Campo Grande, Edif\'{\i}cio C8, P-1749-016 Lisbon, Portugal}
	\affiliation{Instituto de Astrof\'{\i}sica e Ci\^{e}ncias do Espa\c{c}o, Faculdade de
		Ci\^encias da Universidade de Lisboa, Campo Grande, Edif\'{\i}cio C8,
		P-1749-016 Lisbon, Portugal;\\
	}%
	
	\author{Salvatore Capozziello}
	\email{capozziello@na.infn.it}
	\affiliation{Dipartimento di Fisica ``E. Pancini", Universit\`a di Napoli ``Federico II", Complesso Universitario di Monte Sant’ Angelo, Edificio G, Via Cinthia, I-80126, Napoli, Italy,}
	\affiliation{Istituto Nazionale di Fisica Nucleare (INFN), sez. di Napoli, Via Cinthia 9, I-80126 Napoli, Italy,}
	\affiliation{Scuola Superiore Meridionale, Largo S. Marcellino, I-80138, Napoli, Italy.}
	
	\author{Francisco S.N Lobo}%
	\email{fslobo@fc.ul.pt}
	\affiliation{%
		Departamento de F\'{i}sica, Faculdade de Ci\^{e}ncias da Universidade de Lisboa, Campo Grande, Edif\'{\i}cio C8, P-1749-016 Lisbon, Portugal}
	\affiliation{Instituto de Astrof\'{\i}sica e Ci\^{e}ncias do Espa\c{c}o, Faculdade de
		Ci\^encias da Universidade de Lisboa, Campo Grande, Edif\'{\i}cio C8,
		P-1749-016 Lisbon, Portugal;\\
	}%
	
	\author{José Pedro Mimoso}%
	\email{jpmimoso@fc.ul.pt}
	\affiliation{%
		Departamento de F\'{i}sica, Faculdade de Ci\^{e}ncias da Universidade de Lisboa, Campo Grande, Edif\'{\i}cio C8, P-1749-016 Lisbon, Portugal}
	\affiliation{Instituto de Astrof\'{\i}sica e Ci\^{e}ncias do Espa\c{c}o, Faculdade de
		Ci\^encias da Universidade de Lisboa, Campo Grande, Edif\'{\i}cio C8,
		P-1749-016 Lisbon, Portugal;\\
	}%

\date{\today}

\begin{abstract}
Hybrid metric-Palatini gravity unifies the metric and Palatini formalisms while preserving a propagating scalar degree of freedom, offering a compelling route to modified gravity consistent with current observations. Motivated by this success, we consider an extended framework -- the hybrid metric-Palatini scalar-tensor (HMPST) theory -- in which an additional scalar field $\phi$ modulates the curvature couplings, enriching the dynamics and enabling nontrivial self-interactions through scalar potentials. We focus on the analytically tractable linear-$f(\hat{R})$ subclass and study its cosmological, strong-field, and weak-field regimes. In homogeneous and isotropic settings, we identify de Sitter and matter-dominated cosmological solutions describing accelerated expansion and early-universe behavior. For static, spherically symmetric configurations, the field equations yield analytic solutions generalizing the Janis-Newman-Winicour and Buchdahl metrics, including the Schwarzschild-de Sitter limit. In the weak-field regime, linearized perturbations around Minkowski space lead to Yukawa-type corrections to the gravitational potential, with an effective Newton constant $G_{\rm eff}$ and post-Newtonian parameter $\gamma$ that recover General Relativity for heavy or weakly coupled scalars. These results show that the linear-$f(\hat{R})$ HMPST subclass provides a consistent and unified description of gravity across cosmological, astrophysical, and Solar System scales, offering a fertile framework for connecting modified gravity to observations and effective field-theoretic extensions.
\end{abstract}



\maketitle


\section{Introduction}

General Relativity (GR), formulated by Einstein over a century ago~\cite{Einstein:1916vd}, remains a cornerstone of modern theoretical physics. It has successfully passed an extensive range of experimental tests with remarkable precision, particularly in weak-field and highly symmetric regimes~\cite{Will:2014kxa}. Its predictions are consistently confirmed by observations of planetary motion, light deflection, gravitational time dilation, and the dynamics of binary pulsars and compact objects~\cite{Taylor:1982zz,Hulse:1974eb,Kramer:2021jcw,LIGOScientific:2016aoc}. These achievements hold across small and intermediate length scales, from the Solar System to strong-field systems such as neutron stars and black holes.  
Nevertheless, persistent challenges at cosmological scales have prompted a critical reassessment of whether GR provides a complete description of gravitation across all regimes. Observations including the accelerated expansion of the Universe, inferred from Type~Ia supernovae~\cite{Riess:1998cb,Perlmutter:1998np}, and the flat rotation curves of galaxies~\cite{Rubin:1980zd}, suggest the presence of dark energy and dark matter, or alternatively, the need for modifications to the gravitational dynamics themselves.

In light of these challenges, alternative theories of gravity have gained significant attention, particularly those modifying the Einstein–Hilbert action, the cornerstone of GR that is linear in the Ricci scalar $R$. A prominent approach extends this action by introducing curvature-dependent terms, thereby modifying the field equations and potentially revealing new physical phenomena. 
Among these proposals, $f(R)$ gravity stands out for its mathematical simplicity and phenomenological depth. In such models, the gravitational Lagrangian density becomes a non-linear function of the Ricci scalar, $f(R)$, introducing additional dynamical degrees of freedom \cite{Capozziello:2011et, DeFelice:2010aj, Nojiri:2010wj}. The motivations and implications of these extensions have been widely studied \cite{Lobo:2008sg, Avelino:2016lpj, Book, CANTATA:2021ktz, CosmoVerseNetwork:2025alb}. 
These theories recover GR in the weak-field regime while permitting low-curvature deviations that can account for cosmic acceleration \cite{Capozziello:2002rd,Carroll:2003wy} and galactic dynamics  \cite{Capozziello:2006ph,Boehmer:2007kx,Boehmer:2007glt} without invoking dark matter or a cosmological constant \cite{Copeland:2006wr}. Moreover, they naturally emerge in frameworks such as quantum gravity, extra dimensions, and string theory, and can be recast as scalar–tensor models, offering fertile ground for both theoretical exploration and empirical tests.

In the study of gravitational theories derived from generalized curvature functionals, such as $f(R)$ gravity, the choice of the variational principle critically determines the resulting dynamics and physical interpretation. Two main formulations exist: the metric formalism and the Palatini formalism. In the metric formalism \cite{Capozziello:2011et, DeFelice:2010aj}, the action is varied with respect to the metric, assuming the connection to be the Levi-Civita one. This leads to fourth-order field equations, since the non-linear dependence on the Ricci scalar $R$ introduces higher derivatives. Although mathematically consistent, these equations are often difficult to handle, particularly in strong-curvature regimes. In contrast, the Palatini formalism treats the metric and affine connection as independent variables. Applied to $f(R)$ gravity, it yields second-order field equations \cite{Olmo:2011uz}, which are more tractable but feature a connection that depends algebraically on the matter fields, as it couples directly to the stress-energy tensor. Consequently, the affine structure acquires a non-purely geometric character, leading to non-standard matter interactions and a more intricate solution space.

An important distinction between the metric and Palatini formulations of $f(R)$ gravity emerges through their scalar-tensor representations. In the metric approach, the theory can be expressed as a scalar-tensor model with a dynamical scalar field \(\phi = df/dR\) satisfying a second-order differential equation. This scalar introduces an additional gravitational degree of freedom and must be sufficiently light to influence cosmological dynamics, allowing for long-range effects such as cosmic acceleration. However, such light scalars face stringent constraints from laboratory experiments, Solar System tests, and astrophysical observations, which restrict their effective range and coupling unless screening mechanisms (e.g., chameleon or symmetron models) suppress their influence in high-density regions \cite{Joyce:2014kja, Brax:2012bsa, Koivisto:2012za, Brax:2012gr}.

While the metric formalism features a propagating scalar degree of freedom, it struggles to satisfy both cosmological and local gravity constraints simultaneously. In contrast, the Palatini approach replaces this dynamical scalar with an algebraic relation tying $\phi$ directly to the trace of the matter stress-energy tensor, eliminating its propagation. This non-dynamical scalar modifies gravity through a matter-dependent coupling, leading to notable issues: instantaneous response to matter can cause instabilities in cosmological perturbations \cite{Koivisto:2006ie, Koivisto:2005yc} and problematic effects at microscopic scales, where the modified connection conflicts with precision atomic experiments \cite{Olmo:2006zu, Olmo:2008ye}.

To address the shortcomings inherent in both the metric and Palatini formalisms of $f(R)$ gravity, an alternative approach has been proposed that unifies their advantageous features while avoiding their respective drawbacks. This framework, known as \emph{hybrid metric-Palatini gravity}, constructs the total gravitational action by superimposing the conventional Einstein-Hilbert term---constructed from the Ricci scalar $R$ associated with the Levi-Civita connection---with a nontrivial function of a Palatini-type scalar curvature $\hat{R}$, derived from an independent affine connection that is not constrained to be metric-compatible \cite{Harko:2011nh}. The resulting action is given by $S = \int d^4x \sqrt{-g} \left[ R + f(\hat{R}) \right]$,
where $R$ ensures compatibility with well-established gravitational behavior in strong-field and local regimes, and the term $f(\hat{R})$ introduces modifications that become relevant at large distances or in low-curvature environments.

The hybrid metric-Palatini construction effectively combines the key advantages of both the metric and Palatini formulations. It retains the well-tested local behavior of GR through the linear Ricci scalar term $R$, while the additional $f(\hat{R})$ component introduces extra dynamical degrees of freedom capable of explaining cosmic acceleration and other large-scale effects \cite{Harko:2011nh,Capozziello:2012ny,Capozziello:2013uya}. When expressed in its scalar–tensor representation, the theory incorporates a genuine kinetic term for the scalar field, endowing it with true dynamics governed by second-order field equations \cite{Harko:2018ayt}. This property distinguishes it from the purely algebraic scalar field of Palatini $f(R)$ gravity and avoids the higher-order instabilities typical of the metric formulation.

A key virtue of the hybrid approach lies in its natural consistency with local gravitational tests. The scalar degree of freedom emerging in this framework does not necessitate ad hoc screening mechanisms—such as the chameleon or Vainshtein effects—to remain compatible with Solar System and laboratory observations \cite{Capozziello:2015lza,Harko:2018ayt,Harko:2020ibn}. Consequently, hybrid metric-Palatini gravity stands out as a theoretically consistent and phenomenologically viable extension of Einstein’s theory. Its scalar–tensor correspondence enables a systematic analysis of cosmological dynamics, perturbative stability, and structure formation, providing a unified framework that extends the geometric foundations of gravity and offers promising avenues for addressing fundamental problems such as dark energy, inflation, and modified gravity–matter interactions.

Building on the success of hybrid metric–Palatini gravity in merging the metric and Palatini contributions while retaining a viable scalar degree of freedom, it is natural to extend the framework by introducing an explicit scalar field $\phi$ that modulates the curvature couplings. This generalization enhances the model’s flexibility in describing cosmological dynamics, controlling the scalar kinetic structure, and incorporating self-interactions through a potential. It unifies the metric and Palatini limits within a single variationally consistent theory, formulated as a scalar–tensor extension where $\phi$ governs the dynamical sector through a kinetic function $K(\phi)$ and potential $V(\phi)$. Variation with respect to the independent connection enforces a conformal relation between the associated metrics, allowing $\hat{R}$ to depend algebraically on $R$ and $\phi$ and yielding an equivalent scalar–tensor representation with minimal coupling to matter and standard geodesic motion. This analysis will be explored in detail throughout this work.

This paper is organized as follows. In Sec.~\ref{SectionII}, we provide an overview of the (generalized) hybrid metric–Palatini framework, summarizing its theoretical foundations and key motivations. Section~\ref{SectionIII} introduces the generalized hybrid metric–Palatini scalar–tensor theory (HMPST), presenting the general action, deriving the corresponding field equations, and discussing specific functional forms of the scalar couplings. In Sec.~\ref{SectionIV}, we specialize to the analytically tractable linear-$f(\hat{R})$ subclass, analyzing the resulting field equations and dynamical behavior. Section~\ref{SectionV} examines the theory in different conformal frames, clarifying the relationships between the Jordan and Einstein frame representations. The cosmological implications of the HMPST framework are investigated in Sec.~\ref{SectionVI}, where both de~Sitter and matter-dominated scenarios are explored in detail. Section~\ref{SectionVII} focuses on vacuum solutions and their geometrical properties, while Sec.~\ref{SectionVIII} analyzes the weak-field and slow-motion limits, establishing a consistency with Solar System tests. 
In Sec.~\ref{Section:Discussion}, we discuss the physical motivation and interpretation of the HMPST framework, emphasizing the role of the scalar sector, the relevant energy scales, and the recovery of GR.
Finally, Sec.~\ref{Section:Conclusion} summarizes the main results and conclusions of the work, highlighting possible directions for future research.

\section{(Generalized) hybrid metric-Palatini gravity: an overview}
\label{SectionII}

In this section, we present a brief overview of the hybrid metric--Palatini gravity formalism, including the essential definitions. This theory combines the standard Einstein--Hilbert action with a Palatini-type modification involving a function $f(\hat{R})$, where the Palatini curvature scalar is defined as $\hat{R} \equiv g^{\mu\nu}\hat{R}_{\mu\nu}$, and the Palatini Ricci tensor $\hat{R}_{\mu\nu}$ is constructed from an independent connection $\hat{\Gamma}^\alpha_{\mu\nu}$ according to $\hat{R}_{\mu\nu} \equiv \hat{\Gamma}^\alpha_{\mu\nu , \alpha} - \hat{\Gamma}^\alpha_{\mu\alpha , \nu} + \hat{\Gamma}^\alpha_{\alpha\lambda}\hat{\Gamma}^\lambda_{\mu\nu} - \hat{\Gamma}^\alpha_{\mu\lambda}\hat{\Gamma}^\lambda_{\alpha\nu}$. 
The resulting action for hybrid metric--Palatini gravity is then expressed  as~\cite{Capozziello:2015lza,Harko:2011nh}
\begin{align}\label{eq:Shybrid}
	S= &\frac{1}{2\kappa^2}\int_{\mathcal{M}} \mathrm{d}^4 x\ \sqrt{-g} \left[ R + f(\hat{R})\right] + S_m \,,
\end{align}
where $\mathcal{M}$ is a 4-dimensional Lorentzian manifold, $\kappa^2\equiv 8\pi G=M_{Pl}^{-2}$ with $M_{Pl}\simeq2.4\times 10^{18}\ \text{GeV}$ being the reduced Plank mass. $S_m$ is the matter action defined as follows
\begin{equation}
    S_m = \int_{\mathcal{M}} \mathrm{d}^4 x\ \sqrt{-g} \ \mathcal{L}_m({\Psi},g_{\mu\nu}) \,,
\end{equation}
where $\mathcal{L}_m(\Psi, g_{\mu\nu})$ is the matter Lagrangian that depends on the metric $g_{\mu\nu}$ and the matter fields $\Psi$.

Analogously to the procedure in metric $f(R)$ gravity, the geometric formulation of $f(\hat{R})$ can be recast into an equivalent scalar--tensor representation. Introducing an auxiliary field $\chi$, one can express the action as
\begin{equation} \label{eq:S_scalar0}
S= \frac{1}{2\kappa^2}\int_{\mathcal{M}} \mathrm{d}^4 x\ \sqrt{-g} \left[ R + f(\chi)+f_\chi(\hat{R}-\chi)\right] +S_m \,,
\end{equation}
where $f_\chi\equiv df/d\chi$. 
Observing that $\chi = \hat{R}$ recovers the original action, and defining $\varphi \equiv f_\chi$ and $V(\varphi) = \chi f_\chi - f(\chi)$, the action~\eqref{eq:S_scalar0} can be rewritten as
\begin{equation} \label{eq:S_scalar1}
S= \frac{1}{2\kappa^2}\int_{\mathcal{M}} \mathrm{d}^4 x\ \sqrt{-g} \left[ R + \varphi\hat{R}-V(\varphi)\right] +S_m \,.
\end{equation}

Varying the preceding action with respect to the independent connection leads to a field equation showing that it coincides with the Levi-Civita connection of a metric $t_{\mu\nu}$, related to the Einstein-frame metric by the conformal transformation $t_{\mu\nu} = \varphi g_{\mu\nu}$~\cite{Capozziello:2015lza,Harko:2011nh}. Consequently, the Palatini Ricci tensor $\hat{R}_{\mu\nu}$ is related to the metric Ricci tensor $R_{\mu\nu}$ as
\begin{equation}
\hat{R}_{\mu\nu}=R_{\mu\nu}+\frac{3}{2\varphi^2}\partial_\mu \varphi \partial_\nu \varphi-\frac{1}
{\varphi}\left(\nabla_\mu \nabla_\nu \varphi+\frac{1}{2}g_{\mu\nu}\Box\varphi\right) \,,
\end{equation}
where $\nabla_\mu$ denotes the covariant derivative and $\Box \equiv \nabla^\mu \nabla_\mu$ represents the d'Alembert operator. This relation between the two geometric scalars can then be substituted into Eq.~\eqref{eq:S_scalar1} to yield the corresponding scalar-tensor theory, given by
\begin{eqnarray} \label{eq:S_scalar2}
S=\frac{1}{2\kappa^2}\int_{\mathcal{M}} \mathrm{d}^4 x\ \sqrt{-g}\Bigg[ (1+\varphi)R +\frac{3}{2\varphi}\partial_\mu \varphi
\partial^\mu \varphi 
	\nonumber \\
-V(\varphi)\Big]+S_m .
\end{eqnarray}

Shortly after the formulation of hybrid metric-Palatini gravity, a generalization was proposed, characterized by the action~\cite{Tamanini:2013ltp}
\begin{equation}
S= \frac{1}{2\kappa^2}\int_{\mathcal{M}} \mathrm{d}^4 x\ \sqrt{-g}  f(R,\hat{R}) + S_m \,.
\end{equation}
Following the same procedure as in the hybrid metric--Palatini gravity case, this generalized theory can be expressed in scalar--tensor form, yielding
\begin{align}\label{eq: General HMPG}
  S=\frac{1}{2\kappa^2}\int_{\mathcal{M}} d^4x\sqrt{-g} \left[\xi R-\varrho \mathcal{R}- V(\xi,\varrho)\right] \,,
\end{align}
where the scalar fields are defined as $\xi = \frac{\partial f(\alpha, \beta)}{\partial \alpha}$ and $\varrho = -\frac{\partial f(\alpha, \beta)}{\partial \beta}$, and the potential is defined as
\begin{align}\label{eq: Stt General HMPG}
  V(\xi,\varrho)=-f(\alpha(\xi),\beta(\varrho))+\xi\,\alpha(\xi) -\varrho\,\beta(\varrho) \,.
\end{align}

As expected, the field equation for the independent connection establishes the relation $\hat{g}_{\mu\nu} = -\varrho g_{\mu\nu}$, which allows one to rewrite Eq.~\eqref{eq: Stt General HMPG} as~\cite{Tamanini:2013ltp}
\begin{align}
  S=\frac{1}{2\kappa^2}\int_{\mathcal{M}} d^4x\sqrt{-g} \left[\left(\xi-\varrho\right)R-\frac{3}{2\xi}(\partial\varrho)^2 -V(\xi,\varrho)\right] \,.
\end{align}
By defining $\psi = \xi- \varrho$, the above action can be rewritten as~\cite{Tamanini:2013ltp}
\begin{align}
  S=\frac{1}{2\kappa^2}\int_{\mathcal{M}} d^4x\sqrt{-g} \left[\psi\,R-\frac{3}{2\varrho
    }(\partial\varrho)^2 -W(\psi,\varrho)\right] \,,
\end{align}
resulting in an action that corresponds to a Brans--Dicke theory with a vanishing kinetic term and a minimal coupling between the Brans--Dicke scalar and $\varrho$, encoded in the potential $W(\psi, \varrho)$.

\section{Generalized Hybrid Metric-Palatini Scalar-Tensor Theory}\label{SectionIII}

\subsection{Action and field equations}

Motivated by the success of hybrid metric--Palatini gravity in combining metric and Palatini contributions while maintaining a viable scalar degree of freedom, it is natural to consider a further generalization in which an explicit scalar field $\phi$ modulates the curvature couplings. Such a construction allows for greater flexibility in modeling cosmological dynamics, controlling the scalar kinetic term, and encoding self-interactions through a potential. Furthermore, by encompassing both metric and Palatini limits, this approach unifies previously studied theories within a single variationally consistent framework.

The action 
\begin{align}\label{eq:general_hybrid_STT}
	S = &\frac{1}{2\kappa^2}\int_{\mathcal{M}} \mathrm{d}^4 x\ \sqrt{-g} \left[ F\left(\phi,R,\hat{R}\right) - K(\phi) g^{\mu\nu} \partial_\mu \phi \partial_\nu \phi \right.\nonumber\\
	&+ V(\phi) \Big] +  \int_{\mathcal{M}} \mathrm{d}^4 x\ \sqrt{-g} \ \mathcal{L}_m({\Psi},g_{\mu\nu}) \,,
\end{align}
represents a natural scalar--tensor generalization of hybrid metric--Palatini gravity, where $\phi$ is a scalar field, $K(\phi)$ is a general well-behaved function of $\phi$, and $V(\phi)$ is a generic potential for the scalar field. Variation with respect to the independent connection enforces it to be the Levi--Civita connection of a metric conformally related to $g_{\mu\nu}$, ensuring that $\hat{R}$ is algebraically linked to $R$ and $\phi$, which allows the action to be rewritten in an equivalent scalar--tensor form. The kinetic function $K(\phi)$ and potential $V(\phi)$ govern the dynamics and self-interactions of $\phi$, while the minimal matter coupling preserves the standard geodesic motion of test particles. Altogether, Eq.~\eqref{eq:general_hybrid_STT} provides a consistent, variationally well-posed framework that generalizes hybrid metric--Palatini gravity while retaining control over propagating degrees of freedom and reproducing known limits~\cite{Capozziello:2015lza,Harko:2011nh,Tamanini:2013ltp}.

The action \eqref{eq:general_hybrid_STT} depends on three fundamental quantities: the metric $g_{\mu\nu}$, the scalar field $\phi$, and the independent connection $\hat{\Gamma}^\alpha_{\mu\nu}$. As discussed in the previous section, the independent connection is not arbitrary but related to the metric $g_{\mu\nu}$. To make this explicit, we start by varying the action with respect to $\hat{\Gamma}^\alpha_{\mu\nu}$, yielding
\begin{align}
	\delta_{\hat{\Gamma}} S &= \int_{\mathcal{M}} \mathrm{d}^4 x\ \sqrt{-g} \left[ F_{\hat{R}} g^{\mu\nu} \delta \hat{R}_{\mu\nu} \right] \nonumber\\
	&= \int_{\mathcal{M}} \mathrm{d}^4 x\ \sqrt{-g} \left[ F_{\hat{R}} g^{\mu\nu} \left( \hat{\nabla}_\sigma \delta \hat{\Gamma}^\sigma_{\nu\mu} - \hat{\nabla}_\nu \delta \hat{\Gamma}^\sigma_{\sigma\mu} \right) \right] = 0\,,
\end{align}
which can equivalently be written as
\begin{eqnarray}
	&& \delta_{\hat{\Gamma}} S = \int_{\mathcal{M}} \mathrm{d}^4 x\ \Big[ \delta^\nu_\sigma \hat{\nabla}_\lambda \left( \sqrt{-g} F_{\hat{R}} g^{\mu\lambda} \right) 
		\nonumber \\
	&&  \hspace{2cm}  - \hat{\nabla}_\sigma \left( \sqrt{-g} F_{\hat{R}} g^{\mu\nu} \right) \Big] \delta \hat{\Gamma}^\sigma_{\nu\mu} 
		\nonumber\\
	&& \hspace{-0.5cm} + \int_{\mathcal{M}} \mathrm{d}^4 x\ \hat{\nabla}_\sigma \left[ \sqrt{-g} F_{\hat{R}} \left( g^{\mu\nu} \delta \hat{\Gamma}^\sigma_{\nu\mu} - g^{\mu\sigma} \delta \hat{\Gamma}^\lambda_{\lambda\mu} \right) \right] = 0\,,
\end{eqnarray}
where $\hat{\nabla}$ is the covariant derivative constructed from the independent connection and the second term on the RHS is a boundary term.

Applying the principle of stationary action then leads to the field equation
\begin{eqnarray}\label{FE Gamma}
	\hat{\nabla}_\sigma \left( \sqrt{-g} F_{\hat{R}} g^{\mu\nu} \right) - \frac{1}{2} \Big[ \hat{\nabla}_\lambda \left( \sqrt{-g} F_{\hat{R}} g^{\mu\lambda} \right) \delta^\nu_\sigma 
		\nonumber \\
	+ \hat{\nabla}_\lambda \left( \sqrt{-g} F_{\hat{R}} g^{\lambda\nu} \right) \delta^\mu_\sigma \Big] = 0\,.
\end{eqnarray}

By contracting the indices $\sigma$ and $\nu$, this reduces to
\begin{equation}
	\hat{\nabla}_\lambda \left( \sqrt{-g} F_{\hat{R}} g^{\lambda\mu} \right) = 0\,,
\end{equation}
which allows \eqref{FE Gamma} to be rewritten in the simpler form
\begin{equation}
	\hat{\nabla}_\sigma \left( \sqrt{-g} F_{\hat{R}} g^{\mu\nu} \right) = 0\,.
\end{equation}

This equation implies that the independent connection is the Levi--Civita connection of a metric conformally related to the Einstein-frame metric $g_{\mu\nu}$, i.e.,
\begin{equation}\label{eq:conformal indp g and Einst g}
	\hat{g}_{\mu\nu} = F_{\hat{R}} g_{\mu\nu} = \Omega^2 g_{\mu\nu}\,.
\end{equation}

Consequently, the Palatini Ricci tensor can be expressed in terms of the metric Ricci tensor as
\begin{align}\label{eq:R paltini to R}
	\hat{R}_{\mu\nu} &= R_{\mu\nu} + \frac{3}{2(F_{\hat{R}})^2} \partial_\mu F_{\hat{R}} \, \partial_\nu F_{\hat{R}} \nonumber\\
	&\quad - \frac{1}{F_{\hat{R}}} \left( \nabla_\nu \nabla_\mu F_{\hat{R}} + \frac{1}{2} g_{\mu\nu} \Box F_{\hat{R}} \right)\,,
\end{align}
where $F_{\hat{R}} \equiv \partial F(\phi,R,\hat{R})/\partial \hat{R}$ and $\Box \equiv \nabla^\mu \nabla_\mu$.

This result allows to rewrite the Palatini sector of the field equations in terms of the metric Ricci scalar. The variation with respect to the metric $g_{\mu\nu}$ yields
\begin{align}
    &F_R R_{\mu\nu}-\frac{1}{2}g_{\mu\nu}F + \left(g_{\mu\nu}\Box - \nabla_\mu\nabla_\nu\right)F_R + F_{\hat{R}}\hat{R}_{\mu\nu}
    	\nonumber\\
    & \qquad -K(\phi)\left(\partial_\mu\phi \partial_\nu\phi - \frac{1}{2}g_{\mu\nu}\partial_\alpha\phi \partial^\alpha\phi\right) 
    	\nonumber\\
    & \qquad \qquad
    - \frac{1}{2}g_{\mu\nu} V(\phi) = \kappa^2 T_{\mu\nu}\,,
\end{align}
where the stress-energy tensor $T_{\mu\nu}$ is defined in the usual manner
\begin{equation}\label{eq:Stresse-energy tensor}
    T_{\mu\nu} = -\frac{2}{\sqrt{-g}}\frac{\delta (\sqrt{-g}\mathcal{L}_m)}{\delta g^{\mu\nu}}\, .
\end{equation}

Now, using Eq.~\eqref{eq:R paltini to R} the metric field equation can be rewritten as
\begin{eqnarray}
	&&\left(F_R + F_{\hat{R}}\right)R_{\mu\nu}-\frac{1}{2}g_{\mu\nu}F + \left(g_{\mu\nu}\Box - \nabla_\mu\nabla_\nu\right)F_R 
	\nonumber \\ 
	&& \qquad
	+ \frac{3}{2F_{\hat{R}}} \partial_\mu F_{\hat{R}}\;\partial_\nu F_{\hat{R}} 
	- \left(\nabla_\nu \nabla_\mu F_{\hat{R}} + \frac{1}{2}g_{\mu\nu}\Box F_{\hat{R}}\right)
	\nonumber \\ 
	&& \qquad
	-K(\phi)\left(\partial_\mu\phi \, \partial_\nu\phi - \frac{1}{2}g_{\mu\nu}\partial_\alpha \phi \, \partial^\alpha\phi\right) 
	\nonumber \\ 
	&& \qquad
	- \frac{1}{2}g_{\mu\nu} V(\phi) = \kappa^2 T_{\mu\nu}\,.
\end{eqnarray}

Varying the action with respect to the scalar field $\phi$ provides the field equation 
\begin{equation}
    F_\phi + K_\phi \partial_\mu \phi \partial^\mu \phi +2K\Box\phi + V_\phi=0\,.
\end{equation}

\subsection{Specific functional form}

We now focus on the specific functional form
\begin{equation}
	F(\phi,R,\hat{R}) = f(\phi) \left( R + \beta(\phi) f(\hat{R}) \right),
	\label{Specific_form}
\end{equation}
where $f(\phi)$ is a smooth, well-behaved function of the scalar field $\phi$, and $\beta(\phi)$ is a coupling function that introduces a dynamical interaction between the Palatini sector and the scalar field. This choice is motivated by several considerations. 

First, in the limit $\beta(\phi) \to 0$, the theory naturally reduces to a conventional scalar--tensor theory (STT), thereby ensuring consistency with well-established results in both theoretical and phenomenological contexts. This property guarantees that standard scalar--tensor dynamics are recovered in the absence of the Palatini contribution, providing a clear and controlled baseline for analyzing deviations that arise once $\beta(\phi)$ is nonvanishing.

Second, the inclusion of the Palatini term weighted by $\beta(\phi)$ extends the standard STT by incorporating features of hybrid metric--Palatini gravity. In particular, it allows the independent connection to contribute effectively through an additional scalar degree of freedom, which is dynamically coupled to $\phi$. This construction preserves the variational consistency of the theory, as the independent connection can be eliminated algebraically via the field equations, yielding a scalar--tensor representation in which both scalar degrees of freedom and their interactions are manifest.  

Third, the functions $f(\phi)$ and $\beta(\phi)$ offer flexibility in controlling the strength and character of the scalar--curvature couplings. The function $f(\phi)$ modulates the overall effective gravitational coupling, analogous to the role of the Brans--Dicke field in standard STTs, while $\beta(\phi)$ regulates the influence of the Palatini sector on the dynamics. By adjusting these functions, one can explore a wide range of phenomenological scenarios, from modifications relevant at cosmological scales to small deviations in strong-field regimes, all within a single unified framework.  

Finally, this choice of $F(\phi,R,\hat{R})$ provides a natural scalar--tensor generalization of hybrid metric--Palatini gravity, retaining its core features while embedding them into a broader and more flexible theoretical structure. Consequently, as mentioned in the Introduction, we refer to the resulting theory as the \emph{Hybrid Metric–Palatini Scalar–Tensor theory} (HMPST), highlighting both its hybrid nature and its explicit scalar--tensor formulation. This framework thus offers a controlled and consistent avenue to investigate modified gravitational dynamics beyond standard STTs, while maintaining contact with known limits and ensuring the variational well-posedness of the action, as well as enabling systematic comparisons with existing scalar–tensor constructions.

\section{Hybrid Metric-Palatini Scalar-Tensor}\label{SectionIV}

Thus, taking into account the specific functional form (\ref{Specific_form}), the action (\ref{eq:general_hybrid_STT}) takes the following form
\begin{eqnarray}
	S = \frac{1}{2\kappa^2}\int_{\mathcal{M}} \mathrm{d}^4 x\ \sqrt{-g} \bigg[ f(\phi) \left( R + \beta(\phi)f(\hat{R})\right) 
	\nonumber \\ 
	- \frac{\omega(\phi)}{\phi} g^{\mu\nu} \partial_\mu \phi \partial_\nu \phi + V(\phi) \bigg] + S_m \,,
	\label{eq:S hybrid STT}
\end{eqnarray}
where we choose $K(\phi) = \omega(\phi)/\phi$ in order to recover the standard STT of gravity, with the function $\omega(\phi)$ dynamically regulating the coupling strength between the scalar field and the gravitational sector.

At this stage, it is important to emphasize the following points. In~\cite{Borowiec:2020lfx}, a class of STTs was proposed that shares certain structural similarities with the theory we present here. However, our construction differs in several fundamental aspects. Specifically, we consider a Palatini $f(\hat{R})$ sector that is explicitly coupled to the scalar field $\phi$. As will be shown in detail below, this leads to a framework in which the theory effectively contains two interacting scalar fields: the original scalar field $\phi$, which can be interpreted as a Brans--Dicke-type field, and an additional scalar degree of freedom arising from the $f(\hat{R})$ term. Both scalars are coupled to the Ricci curvature and interact dynamically, while the scalar associated with $f(\hat{R})$ naturally possesses a potential term, ensuring nontrivial dynamics.

Consequently, the theory initially exhibits four degrees of freedom: the metric $g_{\mu\nu}$, the scalar field $\phi$, the independent connection $\hat{\Gamma}^{\alpha}_{\mu\nu}$, and the scalar-like degree of freedom originating from the Palatini function $f(\hat{R})$. The interplay between these degrees of freedom is richer than in standard STTs or in the class proposed in~\cite{Borowiec:2020lfx}, as it combines metric, Palatini, and scalar dynamics within a unified variational framework.  

We also note that related constructions exploring the coupling of scalar fields to the Palatini curvature have been considered in cosmological contexts, particularly in the study of inflationary dynamics, as discussed in~\cite{Zhang:2025tpg,Dimopoulos:2022rdp}. These works highlight the versatility of Palatini scalar--tensor models in generating nontrivial potentials and controlled interactions, reinforcing the motivation for the present HMPST framework.

\subsection{Field equations and dynamics}

To construct a consistent scalar--tensor representation of the theory, we start by promoting the Palatini $f(\hat{R})$ sector from its geometric formulation to a scalar--tensor form. This is achieved by introducing an auxiliary scalar field $\varphi$, which captures the dynamics of the Palatini curvature, while preserving variational consistency. In this approach, the original action can be rewritten as
\begin{align}\label{eq:S hybrid STTP}
	S =& \frac{1}{2\kappa^2} \int_{\mathcal{M}} \mathrm{d}^4 x\ \sqrt{-g} \Big[  f(\phi) \left( R + \beta(\phi) \varphi \hat{R} \right) 
		\nonumber\\
	&- \frac{\omega(\phi)}{\phi} g^{\mu\nu} \partial_\mu \phi \partial_\nu \phi  + V(\phi) - f(\phi) \beta(\phi) U(\varphi) \Big] + S_m \,.
\end{align}

The choice of this functional form is motivated by several considerations. First, the factor $f(\phi)$ allows the scalar field $\phi$ to modulate the overall gravitational coupling, analogous to the Brans--Dicke scalar in standard scalar--tensor theories, while $\omega(\phi)$ determines the kinetic structure and dynamically regulates the interaction strength between $\phi$ and gravity. Second, the coupling function $\beta(\phi)$ links the auxiliary scalar $\varphi$, arising from the Palatini sector, to $\phi$, enabling a controlled hybrid interaction between the metric and Palatini contributions. The potential $U(\varphi)$ ensures nontrivial dynamics for the scalar field associated with the Palatini term, while $V(\phi)$ encodes self-interactions for the original scalar field.

This construction preserves the variational consistency of the theory. Variation with respect to the independent connection enforces it to be the Levi--Civita connection of a metric conformally related to $g_{\mu\nu}$, allowing the elimination of $\hat{R}$ in favor of the scalar field $\varphi$, as discussed in detail in previous sections. Consequently, the action~\eqref{eq:S hybrid STTP} explicitly exhibits the two scalar degrees of freedom, both dynamically coupled to the Ricci curvature and interacting through the potentials $V(\phi)$ and $U(\varphi)$. This ensures that the theory is a well-defined scalar--tensor generalization of hybrid metric--Palatini gravity, with controlled interactions, recoverable limits (for $\beta(\phi) \to 0$ or $U(\varphi) = 0$), and a consistent variational structure.

The action~\eqref{eq:S hybrid STTP} depends on four fundamental quantities: the metric $g_{\mu\nu}$, the scalar field $\phi$, the auxiliary scalar $\varphi$, and the independent connection $\hat{\Gamma}^\alpha_{\mu\nu}$. As before, the field equation for the independent connection can be used to eliminate its explicit dependence from the action. For the specific choice $F(\phi,R,\hat{R}) = f(\phi) \left(R + \beta(\phi) \varphi \hat{R}\right)$, Eq.~\eqref{eq:conformal indp g and Einst g} implies that the connection is compatible with the conformally related metric
\begin{equation}
	\hat{g}_{\mu\nu} = f(\phi)\beta(\phi)\,\varphi\, g_{\mu\nu} = \Omega^2 g_{\mu\nu}\,,
\end{equation}
which in turn allows one to express the Palatini Ricci tensor in terms of the metric Ricci tensor as
\begin{align}
	\hat{R}_{\mu\nu} &= R_{\mu\nu} + \frac{3}{2(f \beta \varphi)^2} \partial_\mu(f \beta \varphi) \partial_\nu(f \beta \varphi) \nonumber\\
	&\quad - \frac{1}{f \beta \varphi} \left( \nabla_\nu \nabla_\mu(f \beta \varphi) + \frac{1}{2} g_{\mu\nu} \Box(f \beta \varphi) \right)\,,
\end{align}
leading to the corresponding Ricci scalar
\begin{align}\label{eq:Ricc P vs Ricc E}
	\hat{R} &= R + \frac{3}{2} h(\phi)^2 \, \partial_\mu \phi \, \partial^\mu \phi + \frac{3}{2\varphi^2} \, \partial_\mu \varphi \, \partial^\mu \varphi \nonumber\\
	&\quad + 3\, h(\phi) \frac{\partial_\mu \phi \, \partial^\mu \varphi}{\varphi} - 3\, h(\phi)\, \Box \phi - 3\, \frac{\Box \varphi}{\varphi}\,,
\end{align}
where the function $h(\phi)$ is defined as
\begin{equation}\label{eq:h function}
	h(\phi) = \frac{f_\phi}{f} + \frac{\beta_\phi}{\beta}\,.
\end{equation}

This representation explicitly shows how the auxiliary scalar $\varphi$ and the original scalar $\phi$ contribute to the effective curvature in the theory, illustrating the dynamical interplay between the two scalar degrees of freedom and the metric. In particular, it highlights the role of the functions $f(\phi)$ and $\beta(\phi)$ in modulating both the kinetic cross-term and the coupling to the Palatini sector.

Substituting the expression for $\hat{R}$ from Eq.~\eqref{eq:Ricc P vs Ricc E} into the action~\eqref{eq:S hybrid STTP} leads to the scalar--tensor form
\begin{eqnarray}\label{eq:Action final crossed}
	S &=& \frac{1}{2\kappa^2}\int_{\mathcal{M}} \mathrm{d}^4x \sqrt{-g} \Big[\left(f(\phi)+f(\phi)\beta(\phi)\varphi\right)R
	\nonumber \\
	&&-\left\{\frac{\omega(\phi)}{\phi}+\varphi\left(\frac{3}{2}\Bar{h}(\phi)h(\phi)+\Bar{h}(\phi)_\phi\right)\right\}\partial_\mu \phi \partial^\mu \phi 
	\nonumber \\
	&& + V(\phi)  +\frac{3}{2}f(\phi)\beta(\phi)\frac{\partial_\mu\varphi\partial^\mu\varphi}{\varphi}-3\Bar{h}(\phi)\partial_\mu\phi\partial^\mu\varphi
	\nonumber \\
	&& -f(\phi)\beta(\phi)U(\varphi)\Big]+S_m\,,
\end{eqnarray}
where we have defined $\bar{h}(\phi) \equiv f(\phi) \beta(\phi) h(\phi)$.

A notable feature of this construction is the presence of two scalar fields, $\phi$ and $\varphi$, both non-minimally coupled to the Ricci scalar, giving the theory a dual scalar--tensor character. This structure may have interesting implications in cosmology, for instance in the context of inflationary dynamics, where the interplay of multiple scalar degrees of freedom can generate nontrivial potentials and interactions. From a theoretical perspective, this framework inherits all the positive features of hybrid metric--Palatini gravity (HMPG), while benefiting from the additional flexibility and rich dynamics of scalar--tensor theories. In particular, the effective gravitational coupling,
\begin{equation}
	G_{\rm eff}^{-1} \sim f(\phi) + f(\phi) \beta(\phi) \varphi,
\end{equation}
is dynamically determined by the interaction between the two scalar fields, with its functional form controlled by the choice of $f(\phi)$ and $\beta(\phi)$, thus allowing for a rich variety of gravitational phenomenology.

Now, varying the action~\eqref{eq:Action final crossed} with respect to the metric $g_{\mu\nu}$ yields the generalized Einstein equations
\begin{align}\label{eq:EF geral metric}
	& F(\phi,\varphi) G_{\mu\nu} + g_{\mu\nu} \Box F(\phi,\varphi) - \nabla_\mu \nabla_\nu F(\phi,\varphi) \nonumber\\
	& \quad + \bar{\omega}(\phi,\varphi) \left( \frac{1}{2} g_{\mu\nu} \partial_\alpha \phi \, \partial^\alpha \phi - \partial_\mu \phi \, \partial_\nu \phi \right) - \frac{1}{2} g_{\mu\nu} V(\phi) \nonumber\\
	& \quad + \frac{3 f \beta}{2\varphi} \left( \partial_\mu \varphi \, \partial_\nu \varphi - \frac{1}{2} g_{\mu\nu} \partial_\alpha \varphi \, \partial^\alpha \varphi \right) + \frac{1}{2} f \beta g_{\mu\nu} U(\varphi) \nonumber\\
	& \quad  + 3 \bar{h} \left( g_{\mu\nu} \partial_\alpha \phi \, \partial^\alpha \varphi - \partial_\mu \phi \, \partial_\nu \varphi \right) = \kappa^2 T_{\mu\nu}\,.
\end{align} 

Variation with respect to the scalar field $\phi$ gives the corresponding field equation
\begin{align}\label{eq:EF geral phi}
	F_\phi R &+ \bar{\omega}_\phi(\phi,\varphi) \, \partial_\mu \phi \, \partial^\mu \phi + 2 \bar{\omega}(\phi,\varphi) \Box \phi + V_\phi \nonumber\\
	&+ \frac{3 (f\beta)_\phi}{2\varphi} \, \partial_\mu \varphi \, \partial^\mu \varphi - 3 \bar{h}_\phi \, \partial_\mu \phi \, \partial^\mu \varphi + 3 \partial_\mu \bar{h} \, \partial^\mu \varphi \nonumber\\
	&+ 3 \bar{h} \Box \varphi - (f \beta)_\phi \, U(\varphi) = 0\,,
\end{align}
while variation with respect to the auxiliary scalar $\varphi$ leads to
\begin{align}\label{eq:EF geral varphi}
	F_\varphi R &- \bar{\omega}_\varphi \, \partial_\mu \phi \, \partial^\mu \phi +\frac{3f\beta}{2\varphi^2} \partial_\mu \varphi \partial^\mu \varphi - \frac{3 f\beta}{\varphi} \partial_\mu \phi \partial^\mu \varphi\nonumber\\
	&- \frac{3f\beta}{\varphi} \Box\varphi + 3 \partial_\mu \bar{h} \, \partial^\mu \phi + 3 \bar{h} \Box \phi - f \beta U_\varphi = 0\,.
\end{align}

For notational convenience, we have introduced
\begin{equation}
	F(\phi,\varphi) \equiv f(\phi) \left[ 1 + \beta(\phi) \varphi \right]\,,
\end{equation}
and
\begin{equation}
	\bar{\omega}(\phi,\varphi) \equiv \frac{\omega(\phi)}{\phi} + \varphi \left( \frac{3}{2} \bar{h}(\phi) h(\phi) + \bar{h}_\phi(\phi) \right)\,,
\end{equation}
where as before $\bar{h}(\phi) = f(\phi) \beta(\phi) h(\phi)$ and $h(\phi)$ is defined in Eq.~\eqref{eq:h function}.

The field equation~\eqref{eq:EF geral metric} can be conveniently rewritten in the form
\begin{equation}
	G_{\mu\nu} = \kappa^2 T_{\mu\nu} + H_{\mu\nu}(\phi,\varphi)\,,
\end{equation} 
where the tensor $H_{\mu\nu}(\phi,\varphi)$ encodes the contributions from the scalar fields and is given by
\begin{eqnarray}
	&&H_{\mu\nu}(\phi,\varphi) = \frac{1}{F}\bigg[\nabla_\mu\nabla_\nu F - g_{\mu\nu} \Box F 
	\nonumber \\
	&& \hspace{0.55cm} + \Bar{\omega}(\phi,\varphi) \left(\frac{1}{2}g_{\mu\nu} \partial_\alpha\phi\partial^\alpha\phi +\partial_\mu\phi\partial_\nu\phi\right)
	+\frac{1}{2}g_{\mu\nu}V(\phi) 
	\nonumber \\ 
	&& \hspace{0.55cm} -\frac{3f\beta}{2\varphi}\left( \partial_\mu\varphi\partial_\nu\varphi+\frac{1}{2}g_{\mu\nu} \partial_\alpha\varphi\partial^\alpha\varphi\right) 
	\nonumber \\ 
	&& \hspace{0.55cm}
	-3\Bar{h}\left(g_{\mu\nu} \partial_\alpha\phi\partial^\alpha\varphi +\partial_\mu\phi\partial_\nu\varphi\right)-\frac{1}{2}f\beta g_{\mu\nu} U(\varphi) \bigg]\,.
\end{eqnarray}
This term can be interpreted as an effective additional contribution to the matter sector of the theory, with the scalar field $\phi$ effectively acting as a dynamical coupling controlling the strength of this new sector.

From the action~\eqref{eq:Action final crossed}, it is evident that the fundamental dynamics of the theory arise from the interplay between the two scalar fields $\phi$ and $\varphi$. In particular, the functions $f(\phi)$ and $\beta(\phi)$ govern both the coupling of the scalars to the Ricci scalar and their mutual interactions. The detailed structure of the scalar-scalar interactions is further regulated by the function $h(\phi)$ defined in Eq.~\eqref{eq:h function}. 

If one desires to eliminate the direct kinetic interaction between $\phi$ and $\varphi$, it is sufficient to impose
\begin{equation}\label{eq:beta constrangido}
	h(\phi) = 0 \quad \Rightarrow \quad \beta(\phi) = \frac{\lambda}{f(\phi)}\,,
\end{equation}
with $\lambda$ being a constant. Under this condition, the action reduces to the simplified form
\begin{eqnarray}\label{eq:HMPST Action final}
	&& S = \frac{1}{2\kappa^2}\int_{\mathcal{M}} \mathrm{d}^4x \sqrt{-g} \bigg[\left(f(\phi)+\lambda\varphi\right)R-\frac{\omega(\phi)}{\phi}\partial_\mu \phi \, \partial^\mu \phi 
	\nonumber\\
	&& 	\hspace{1cm} +V(\phi)+\frac{3\lambda}{2}\frac{\partial_\mu\varphi\partial^\mu\varphi}{\varphi}-\lambda U(\varphi)\bigg]+S_m\,,
\end{eqnarray}
which preserves the dual scalar-tensor nature while eliminating the kinetic cross-term between the two scalar fields. This form highlights the role of $\phi$ and $\varphi$ in determining the effective gravitational coupling and the dynamics of the scalar sector in a transparent and controlled manner.

Although in the action~\eqref{eq:HMPST Action final} the kinetic interaction between the two scalar fields has been eliminated, their coupling to the Ricci scalar persists, emphasizing the nontrivial interplay between the scalar sector and the metric. Consequently, the essential dynamics of the theory are determined by the function $f(\phi)$ and its interaction with the auxiliary scalar $\varphi$. 

For instance, in the extreme case $f(\phi) = 1$, the coupling between $\phi$ and $R$ is removed, leaving only the coupling between the metric and $\varphi$. This coupling is a direct consequence of the scalar-tensor representation of $f(\hat{R})$ and cannot be eliminated. In this scenario, the kinetic term and potential of $\phi$ remain, and can be interpreted as contributing to the matter sector, potentially acting as an inflaton field.

Another illustrative case occurs when the Palatini sector is linear in $\hat{R}$, i.e. $f(\hat{R}) = A \hat{R} + B$, with $A$ and $B$ constants. Here, the auxiliary scalar and its potential are constant, $\varphi = A \equiv \varphi_0$ and $U(\varphi) = B \equiv U_0$, leading to the action
\begin{eqnarray}\label{eq:linear palatini action}
	S &=& \frac{1}{2\kappa^2}\int_{\mathcal{M}} \mathrm{d}^4x \sqrt{-g} \Bigg[\left(f(\phi)+\lambda\varphi_0\right)R
	\nonumber \\
	&& -\lambda U_0-\frac{\omega(\phi)}{\phi}\partial_\mu \phi \partial^\mu \phi + V(\phi)\Bigg]+S_m\,.
\end{eqnarray}

In this limit, the theory reduces to a conventional scalar-tensor model with an effective cosmological constant $\Lambda_{\rm eff} = -\lambda U_0$, while the effective gravitational coupling receives an additional contribution $\lambda \varphi_0$ alongside the dynamics determined by $f(\phi)$. This illustrates how different choices for $f(\phi)$ and the Palatini sector can lead to qualitatively distinct gravitational behaviors, including scenarios relevant for cosmology and inflation.

For a general choice of $f(\phi)$ and $f(\hat{R})$, the field equations of the theory can be derived either by varying the action~\eqref{eq:HMPST Action final} directly, or equivalently by taking Eqs.~\eqref{eq:EF geral metric}--\eqref{eq:EF geral varphi} with $\bar{h}=0$, which corresponds to imposing the condition~\eqref{eq:beta constrangido}. This yields the metric field equation
\begin{align}\label{eq:FE special case metric}
	&(f(\phi) + \lambda \varphi) G_{\mu\nu} + (g_{\mu\nu} \Box - \nabla_\mu \nabla_\nu) f(\phi) 
	\nonumber\\
	& \quad
	+ \lambda (g_{\mu\nu} \Box - \nabla_\mu \nabla_\nu) \varphi - \frac{1}{2} g_{\mu\nu} V(\phi) + \frac{\lambda}{2} g_{\mu\nu} U(\varphi) \nonumber\\
	& \quad - \frac{\omega}{\phi} \left( \partial_\mu \phi \partial_\nu \phi -\frac{1}{2} g_{\mu\nu} \partial_\alpha \phi \partial^\alpha \phi \right) \nonumber\\
	& \quad + \frac{3 \lambda}{2 \varphi} \left( \partial_\mu \varphi \partial_\nu \varphi - \frac{1}{2} g_{\mu\nu} \partial_\alpha \varphi \partial^\alpha \varphi \right) = \kappa^2 T_{\mu\nu} \,,
\end{align}
while the scalar field equations read
\begin{align}
	f_\phi R &+ \left( \frac{\omega_\phi}{\phi} - \frac{\omega}{\phi^2} \right) \partial_\mu \phi \partial^\mu \phi + 2 \frac{\omega}{\phi} \Box \phi + V_\phi = 0 \,, \label{eq:FE special case phi} \\
	R &+ \frac{3}{2} \frac{\partial_\mu \varphi \partial^\mu \varphi}{\varphi^2} - 3 \frac{\Box \varphi}{\varphi} - U_\varphi = 0 \,. \label{eq:FE special case varphi}
\end{align}

Taking the trace of the metric field equation~\eqref{eq:FE special case metric} gives
\begin{align}\label{eq:trace metric}
	-(f + \lambda \varphi) R &+ 3 \Box f + 3 \lambda \Box \varphi - 2 V(\phi) + 2 \lambda U(\varphi) 
	\nonumber\\
	&
	+ \omega \frac{\partial_\alpha \phi \partial^\alpha \phi}{\phi} - \frac{3 \lambda}{2} \frac{\partial_\alpha \varphi \partial^\alpha \varphi}{\varphi} = \kappa^2 T\,,
\end{align}
where $T$ denotes the trace of the matter stress-energy tensor.

Combining Eqs.~\eqref{eq:FE special case phi} and~\eqref{eq:FE special case varphi}, one can express the dynamics of $\varphi$ in terms of $\phi$ as
\begin{eqnarray}
	&&-\frac{3}{2} \frac{\partial_\mu \varphi \partial^\mu \varphi}{\varphi^2} +3\frac{\Box\varphi}{\varphi} + U_\varphi= 
	\nonumber \\
	&& \qquad \frac{1}{f_\phi} \left[\left( \frac{\omega}{\phi^2} -\frac{\omega_\phi}{\phi} 
	\right)\partial_\mu \phi \partial^\mu \phi - 2\frac{\omega}{\phi} \Box\phi -V_\phi\right]\,,
\end{eqnarray}
which demonstrates that, although there is no explicit kinetic coupling between $\varphi$ and $\phi$, they remain dynamically linked through their gravitational interactions.

Finally, taking the covariant derivative of Eq.~\eqref{eq:FE special case metric} leads to the standard conservation law for the matter stress-energy tensor, $\nabla^\mu T_{\mu\nu} = 0 $, ensuring the consistency of the theory with the diffeomorphism invariance of the action.

\section{Conformal frames}\label{SectionV}

In any scalar-tensor theory formulated in the Jordan frame, where the action is initially defined with a non-minimal coupling between the scalar fields and the Ricci scalar, it is possible to perform a conformal transformation of the metric
\begin{equation}
	g_{\mu\nu} \ \longrightarrow \ \tilde{g}_{\mu\nu} = \Omega^2 g_{\mu\nu}\,,
\end{equation}
which systematically removes the non-minimal coupling. This procedure recasts the gravitational sector into a form resembling the standard Einstein-Hilbert action, while the scalar fields remain, now minimally coupled to the curvature but generally acquiring a modified kinetic structure and potential. This formulation is referred to as the Einstein frame. 

For the HMPST action~\eqref{eq:HMPST Action final}, the appropriate conformal factor that eliminates the non-minimal coupling between the scalar fields and the Ricci scalar is
\begin{equation}
	\tilde{g}_{\mu\nu} = \left(f(\phi) + \lambda \varphi \right) g_{\mu\nu}\,.
\end{equation}

In terms of this new metric and its associated scalar curvature $\tilde{R}$, the action can be expressed as
\begin{align}
	S = \frac{1}{2\kappa^2} \int_{\mathcal{M}} \mathrm{d}^4 x \, \sqrt{-\tilde{g}} \Big[ & \tilde{R} + \tilde{\omega}_{AB} \tilde{g}^{\mu\nu} \partial_\mu \psi^A \partial_\nu \psi^B 
	+ \tilde{V} + \tilde{U} \Big] \nonumber\\
	&+ \int_{\mathcal{M}} \mathrm{d}^4 x \, \sqrt{-\tilde{g}} \, \tilde{\mathcal{L}}_m \,,
\end{align}
where $\psi^A$ denotes the scalar fields $(\phi, \varphi)$ with $A = 1,2$, and we have considered the following definitions, for notational simplicity
\begin{align}
	\tilde{\mathcal{L}}_m &= \frac{\mathcal{L}_m}{(f + \lambda \varphi)^2} \,, \\
	\tilde{V} &= \frac{V(\phi)}{(f + \lambda \varphi)^2} \,, \\
	\tilde{U} &= - \frac{\lambda U(\varphi)}{(f + \lambda \varphi)^2} \,.
\end{align}

The kinetic structure of the scalar fields is encoded in the symmetric tensor $\tilde{\omega}_{AB}$, which governs the internal configuration of the scalar sector:
\begin{align}
	\tilde{\omega}_{11} &= - \left( \frac{3}{2} \frac{f'^2}{(f + \lambda \varphi)^2} + \frac{\omega(\phi)}{\phi} \right), \nonumber\\
	\tilde{\omega}_{22} &= - \left( \frac{3}{2} \frac{\lambda^2}{(f + \lambda \varphi)^2} - \frac{3}{2 \varphi} \right), \nonumber\\
	\tilde{\omega}_{12} &= \tilde{\omega}_{21} = \frac{3}{2} \frac{\lambda f'}{(f + \lambda \varphi)} \,, \nonumber
\end{align}
where $f' = \mathrm{d}f/\mathrm{d}\phi$. 

This transformation provides a clear and controlled framework for studying the dynamics of the HMPST theory, as the gravitational sector now follows the standard Einstein-Hilbert form, while all nontrivial interactions and couplings are encoded in the scalar sector. It also facilitates the analysis of cosmological and inflationary scenarios, as the scalar degrees of freedom are now minimally coupled to the curvature but retain a rich kinetic and potential structure.

As a consequence of the conformal transformation, in addition to the usual non-minimal coupling between the scalar fields and the matter sector, the action also develops mixed kinetic terms between the scalar fields. Varying the Einstein-frame action with respect to the metric $\tilde{g}^{\mu\nu}$ yields the gravitational field equations
\begin{align}
	\tilde{G}_{\mu\nu} &+ \tilde{\omega}_{AB} \left( \partial_\mu \psi^A \partial_\nu \psi^B - \frac{1}{2} \tilde{g}_{\mu\nu} \tilde{g}^{\alpha\beta} \partial_\alpha \psi^A \partial_\beta \psi^B \right) \nonumber\\
	&- \frac{1}{2} \tilde{g}_{\mu\nu} (\tilde{V} + \tilde{U}) = \kappa^2 \tilde{T}_{\mu\nu} \,,
\end{align}
where the matter stress-energy tensor in the Einstein frame is defined as
\begin{equation}\label{eq:Stresse-energy tensor2}
	\tilde{T}_{\mu\nu} = - \frac{2}{\sqrt{-\tilde{g}}} \frac{\delta (\sqrt{-\tilde{g}} \tilde{\mathcal{L}}_m)}{\delta \tilde{g}^{\mu\nu}} \,.
\end{equation}

Similarly, variation with respect to the scalar fields $\phi$ and $\varphi$ produces their equations of motion. For $\phi$, one obtains
\begin{align}
	(\tilde{\omega}_{AB})_\phi \, \partial_\mu \psi^A \, \partial_\nu \psi^B 
	- \tilde{\nabla}_\mu (\tilde{\omega}_{11} \partial^\mu \phi) 
	- \tilde{\nabla}_\mu (\tilde{\omega}_{12} \partial^\mu \varphi)
		\nonumber \\
	= - \frac{\tilde{V}_\phi}{2} - \frac{\tilde{U}_\phi}{2} - \kappa^2 (\tilde{\mathcal{L}}_m)_\phi \,,
\end{align}
while for $\varphi$ the equation reads
\begin{align}
	(\tilde{\omega}_{AB})_\varphi \, \partial_\mu \psi^A \, \partial_\nu \psi^B 
	- \tilde{\nabla}_\mu (\tilde{\omega}_{22} \partial^\mu \varphi) 
	- \tilde{\nabla}_\mu (\tilde{\omega}_{12} \partial^\mu \phi)
		\nonumber \\
	= - \frac{\tilde{V}_\varphi}{2} - \frac{\tilde{U}_\varphi}{2} - \kappa^2 (\tilde{\mathcal{L}}_m)_\varphi \,.
\end{align}

These equations explicitly show how the kinetic coupling tensor $\tilde{\omega}_{AB}$ encodes both self-interactions and mixed interactions between the scalar fields in the Einstein frame, while the potentials $\tilde{V}$ and $\tilde{U}$, together with the matter Lagrangian, contribute to the sources driving the dynamics of $\phi$ and $\varphi$. This formulation is particularly convenient for analyzing the cosmological evolution, scalar field stability, and the energy exchange between the gravitational and scalar sectors.

\section{Cosmology in HMPST}\label{SectionVI}

In order to investigate the cosmological implications of the present theory, we consider a homogeneous and isotropic background geometry described by the flat Friedmann-Lemaître-Robertson-Walker (FLRW) metric, which is consistent with current cosmological observations indicating spatial flatness at large scales. The metric is given by
\begin{equation}\label{eq: FLRW metric}
	\mathrm{d}s^{2} = -\mathrm{d}t^{2} + a^{2}(t)\,\mathrm{d}V^{2} \,,
\end{equation}
where $t$ represents the cosmic (comoving) proper time, $a(t)$ is the scale factor characterizing the expansion of the universe, and $\mathrm{d}V$ denotes the spatial line element in comoving coordinates.

For the matter sector, we adopt the standard perfect fluid description, which provides a suitable macroscopic characterization of the cosmic content in terms of an effective energy density and pressure. The corresponding energy--momentum tensor takes the form
\begin{equation}\label{eq:EM tensor}
	T_{\mu\nu} = (\rho + p) u_\mu u_\nu + p\, g_{\mu\nu} \,,
\end{equation}
where $\rho$ and $p$ denote the total energy density and isotropic pressure of the fluid, respectively, while $u^{\mu}$ is the four-velocity of the comoving observer, satisfying the normalization condition $u_{\mu} u^{\mu} = -1$.

Substituting the FLRW metric \eqref{eq: FLRW metric} into the field equations \eqref{eq:FE special case metric}--\eqref{eq:FE special case varphi}, together with the perfect fluid form of the energy-momentum tensor \eqref{eq:EM tensor}, yields the following set of cosmological equations governing the background dynamics of the theory:
\begin{align}\label{eq:00 component}
	3H^{2}\left(f(\phi)+\lambda\varphi\right) + 3H\left(\dot{f} + \lambda\dot{\varphi}\right) 
	+ \lambda\frac{3}{4}\frac{\dot{\varphi}^{2}}{\varphi} 
		\nonumber \\
	- \frac{\omega}{2}\frac{\dot{\phi}^{2}}{\phi}
	+ \frac{1}{2}V - \frac{\lambda}{2}U
	= \kappa^{2}\rho\,,
\end{align}
\begin{align}\label{eq:ii component}
	&-(f+\lambda\varphi)\left(2\dot{H} + 3H^{2}\right) 
	- \left(\Ddot{f} + \lambda\Ddot{\varphi}\right)
	- 3H\left(\dot{f} + \lambda\dot{\varphi}\right)
		\nonumber \\
	& \qquad \qquad - \frac{\omega}{2}\frac{\dot{\phi}^{2}}{\phi} 
	+ \lambda\frac{3}{4}\frac{\dot{\varphi}^{2}}{\varphi}
	- \frac{1}{2}V + \frac{\lambda}{2}U
	= \kappa^{2}w\rho\,,
\end{align}
\begin{equation}\label{eq:phi FLRW}
	6f_{\phi}\left(\dot{H} + 2H^{2}\right)
	-\left(\frac{\omega_{\phi}}{\phi} - \frac{\omega}{\phi^{2}}\right)\dot{\phi}^{2}
	- 2\omega\frac{\Ddot{\phi}}{\phi}
	- 6\omega H\frac{\dot{\phi}}{\phi}
	+ V_{\phi} = 0\,,
\end{equation}
\begin{equation}\label{eq:varphi FRLW}
	6\left(\dot{H} + 2H^{2}\right)
	-\frac{3}{2}\left(\frac{\dot{\varphi}}{\varphi}\right)^{2}
	+ 9H\frac{\dot{\varphi}}{\varphi}
	+ 3\frac{\Ddot{\varphi}}{\varphi}
	- U_{\varphi} = 0\,,
\end{equation}
\begin{equation}\label{eq:continuity}
	\dot{\rho} + 3\frac{\dot{a}}{a}(1+w)\rho = 0\,,
\end{equation}
where we have adopted the barotropic equation of state $p = w\rho$ with $w$ constant. Equation \eqref{eq:00 component} corresponds to the temporal ($00$) component of the metric field equations, while Eq.~\eqref{eq:ii component} represents the spatial ($ii$) component. The scalar field equations for $\phi(t)$ and $\varphi(t)$ are given by Eqs.~\eqref{eq:phi FLRW} and \eqref{eq:varphi FRLW}, respectively. Finally, Eq.~\eqref{eq:continuity} follows from the covariant conservation of the energy–momentum tensor, $\nabla_\mu T^{\mu\nu} = 0$, which remains valid in this framework.

\vspace{0.2cm}
An inspection of these equations reveals that both scalar fields, $\phi(t)$ and $\varphi(t)$, evolve under the influence of the Hubble parameter $H(t)$ and are intricately coupled through the functions $f(\phi)$, $V(\phi)$, and $U(\varphi)$. The structure of the differential equations indicates that analytical solutions can only be obtained for specific choices of these functional forms. In particular, the potential $U(\varphi)$ plays a dual role: it not only governs the dynamics of $\varphi(t)$ but also determines, through the scalar–tensor correspondence, the functional dependence of $f(\hat{R})$ in the Palatini sector. Consequently, special attention will be given to the analysis of $U(\varphi)$, as it directly constrains the class of $f(\hat{R})$ models that admit analytical cosmological solutions. 

To this end, in the following sections we investigate how different assumptions for the scale factor $a(t)$ and the scalar field $\varphi(t)$ influence the form of $U_\varphi$ and, consequently, the reconstruction of the underlying function $f(\hat{R})$. This approach provides a systematic way to explore viable cosmological solutions and to assess their consistency with observational and theoretical expectations.

\subsection{De Sitter case}

We begin our analysis by considering the simplest cosmological scenario -- the de Sitter case -- characterized by a constant Hubble parameter, $H = H_0$. In this regime, the scale factor evolves exponentially as $a(t) = a_0 e^{H_0 t}$, corresponding to a spacetime with constant curvature and vacuum-dominated expansion. Substituting $H = H_0$ into Eq.~\eqref{eq:varphi FRLW}, the dynamical equation for the Palatini scalar field $\varphi$ reduces to
\begin{equation}\label{eq: Potential form 1}
	U_{\varphi} = 12H_0^2
	-\frac{3}{2}\left(\frac{\dot{\varphi}}{\varphi}\right)^2
	+ 9H_0\frac{\dot{\varphi}}{\varphi}
	+ 3\frac{\Ddot{\varphi}}{\varphi}\,.
\end{equation}

To express the potential $U(\varphi)$ explicitly in terms of the scalar field, it is convenient to assume that $\varphi$ can be represented by a smooth, differentiable function of time, $\varphi = g(t)$, where $g(t)$ admits an inverse $t = g^{-1}(\varphi)$. This assumption allows us to reformulate Eq.~\eqref{eq: Potential form 1} as
\begin{align}
	U_{\varphi} &= 12H_0^2
	-\frac{3}{2}\left(\frac{\dot{g}}{g}\right)^2
	+ 9H_0\frac{\dot{g}}{g}
	+ 3\frac{\Ddot{g}}{g} \nonumber\\
	&= 12H_0^2 + \mathbf{G}(t) \nonumber\\
	&= 12H_0^2 + \mathbf{G}\!\left(g^{-1}(\varphi)\right)\,,
\end{align}
where we have defined the auxiliary function $\mathbf{G}(t)$ as
\begin{equation}
	\mathbf{G}(t) \equiv
	-\frac{3}{2}\left(\frac{\dot{g}}{g}\right)^2
	+ 9H_0\frac{\dot{g}}{g}
	+ 3\frac{\Ddot{g}}{g}\,.
\end{equation}

This reformulation provides a convenient method for determining $U(\varphi)$ once a specific functional form for $g(t)$ is chosen. In this approach, the potential $U(\varphi)$ naturally inherits its structure from the time evolution of the scalar field $\varphi(t)$, and thus from the underlying cosmological dynamics encoded in $g(t)$. In particular, different ansätze for $\varphi(t)$ will correspond to distinct families of potentials $U(\varphi)$, each associated with a specific realization of the $f(\hat{R})$ function within the hybrid metric-Palatini framework.

Using now the standard relations between the Palatini curvature potential and the scalar representation,
\begin{equation}
	U_{\varphi} = \hat{R}
	\quad \text{and} \quad
	\varphi \equiv \frac{\mathrm{d}f}{\mathrm{d}\hat{R}}\,,
\end{equation}
we can establish a direct connection between the dynamics of the scalar field and the underlying $f(\hat{R})$ function. Substituting these relations into the previous expression for $U_{\varphi}$ yields the differential equation
\begin{equation}
	12H_0^2 + \mathbf{G}\!\left(g^{-1}\!\left(\frac{\mathrm{d}f}{\mathrm{d}\hat{R}}\right)\right) = \hat{R}\,,
\end{equation}
whose general solution can be written as
\begin{equation}
	f(\hat{R}) = \int \left(g \circ \mathbf{G}^{-1}\right)\!\left(\hat{R} - 12H_0^2\right)\, \mathrm{d}\hat{R} + C\,,
\end{equation}
where $C$ is an integration constant. This result encapsulates the entire class of $f(\hat{R})$ functions that admit analytical scalar-field solutions $\varphi(t)$ in a de Sitter background, characterized by $a(t) \sim e^{H_0 t}$. Thus, given a specific ansatz for $\varphi(t)$—or equivalently for $g(t)$—one can, in principle, reconstruct the corresponding $f(\hat{R})$ function consistent with the assumed cosmological dynamics.

It is important to note that the explicit determination of $\mathbf{G}(t)$ is not strictly necessary. Introducing the auxiliary function $\mathbf{F}(t)$ through the definition $\dot{g}/g = \mathbf{F}(t)$ allows Eq.~\eqref{eq: Potential form 1} to be rewritten in the alternative, yet equivalent, form
\begin{eqnarray}
	\hat{R} &=& 12H_0^2
	+ \frac{3}{2}\!\left[\mathbf{F}\!\left(g^{-1}(\varphi)\right)\right]^2
		\nonumber \\
	&& + 9H_0\,\mathbf{F}\!\left(g^{-1}(\varphi)\right)
	+ 3\dot{\mathbf{F}}\!\left(g^{-1}(\varphi)\right)\,.
\end{eqnarray}
This representation can, in many cases, simplify the reconstruction procedure for $f(\hat{R})$, especially when $\mathbf{F}(t)$ assumes a simple or physically motivated form.\\


To illustrate the method, let us now consider a representative example in which the scalar field follows a power-law evolution of the form
\begin{equation}
	\varphi(t) = A\,t^{\sigma}\,,
\end{equation}
where $A$ and $\sigma$ are constants. Substituting this ansatz into Eq.~\eqref{eq: Potential form 1} and expressing the result in terms of $\varphi$ yields the corresponding Palatini curvature scalar,
\begin{equation}
	\hat{R} = 12H_0^2
	+ 3\sigma\!\left(\frac{\sigma}{2} - 1\right)
	\!\left(\frac{\varphi}{A}\right)^{-\frac{2}{\sigma}}
	+ 9H_0\sigma
	\!\left(\frac{\varphi}{A}\right)^{-\frac{1}{\sigma}}\,.
\end{equation}

It is evident from the previous expression that the explicit functional form of $f(\hat{R})$ depends critically on the parameter $\sigma$, which governs the evolution of the scalar field $\varphi(t) = A t^{\sigma}$. This parameter effectively controls the dynamical coupling between the scalar sector and the geometric curvature, and consequently, the structure of the resulting gravitational Lagrangian. In particular, specific values of $\sigma$ lead to remarkably simple analytical forms for $f(\hat{R})$.

For instance, when $\sigma = 2$, the relation between $\hat{R}$ and $\varphi$ can be inverted and directly integrated to yield
\begin{equation}
	f(\hat{R}) = -\frac{324 H_0^2 A}{-12 H_0^2 + \hat{R}} + C_2\,,
\end{equation}
where $C_2$ denotes an integration constant. This inverse-type functional dependence indicates that in this case, the corresponding $f(\hat{R})$ model resembles a rational deformation of the Einstein-Hilbert action, leading to potentially interesting corrections to GR at high curvature scales.

On the other hand, for $\sigma = -2$, one obtains a markedly different behavior. The integration yields
\begin{align}
	f(\hat{R}) =& \frac{1}{6} \left[15 H_0^2 A \hat{R} + \frac{A \hat{R}^2}{2} \right.\nonumber \\
	&\left. - \sqrt{3}\,H_0 A \left(3 H_0^2 + 2 \hat{R}\right)^{3/2} \right] + C_{-2}\,,
\end{align}
where $C_{-2}$ is an integration constant. This case exhibits a more intricate non-linear dependence on $\hat{R}$, incorporating both polynomial and radical terms, which may introduce richer cosmological dynamics, including possible transitions between effective acceleration regimes or curvature-driven inflationary phases.

Having obtained specific analytical forms of $f(\hat{R})$ that are compatible with the de Sitter background ($H = H_0$), one can subsequently determine the corresponding Palatini curvature $\hat{R}$ using Eq.~\eqref{eq:Ricc P vs Ricc E} with $h(\phi) = 0$. This allows a complete reconstruction of the gravitational dynamics in the hybrid metric–Palatini framework under constant-Hubble expansion.

An interesting observation arises when considering the exponential ansatz for the scalar field, $\varphi(t) = e^{\pm A t}$. Substituting this form into Eq.~\eqref{eq: Potential form 1} leads to
\begin{equation}
	\hat{R} = \text{constant}\,,
\end{equation}
which implies a degenerate situation where the scalar field dynamics effectively vanish, i.e., $\varphi = 0$. Such a result signals a mathematical inconsistency, as the vanishing of $\varphi$ eliminates the scalar degree of freedom introduced in the hybrid framework. Therefore, exponential forms for $\varphi(t)$ are ruled out as viable solutions in the de Sitter scenario, reaffirming that only power-law behaviors (with appropriately chosen $\sigma$) yield consistent and dynamically nontrivial models within this setup.

For the scalar field $\phi$, its evolution is governed not only by the Hubble parameter $H$ but also by the specific functional forms of $f(\phi)$ and $V(\phi)$. Consequently, obtaining analytical solutions for $\phi(t)$ is generally challenging without making simplifying assumptions regarding these functions. Complex or non-linear choices for $f(\phi)$ and $V(\phi)$ often lead to highly non-analytical differential equations.  

To simplify the analysis and gain physical insight, we consider the original Brans–Dicke limit of the theory, characterized by a constant coupling parameter $\omega = \omega_{BD}$ and a vanishing potential, $V(\phi)=0$. Under these assumptions, Eq.~\eqref{eq:phi FLRW} reduces to
\begin{equation}\label{eq:phi BD FLRW}
	6\left(\dot{H} + 2H^2\right)  + \omega_{BD}\frac{\dot{\phi}^2}{\phi^2}  
	- 2\omega_{BD}\frac{\Ddot{\phi}}{\phi} - 6\omega_{BD} H\frac{\dot{\phi}}{\phi} =0\,.
\end{equation}
In the de Sitter case, where $H = H_0$ is constant, we define the auxiliary variable $y \equiv \frac{\dot{\phi}}{\phi}$ to express the equation in a more compact form:
\begin{equation}
	12H_0^2  + \omega_{BD} y^2  - 2\omega_{BD}(y^2+y') - 6\omega_{BD} H_0y =0\,,
\end{equation}
which represents a first-order differential equation for $y(t)$. The general solutions to this equation are given by
\begin{eqnarray}
	y_1 &=& -3H_{0}-\dfrac{\sqrt{-9{H_{0}}^{2}\omega_{BD}-12H_{0}}}{\sqrt{\omega_{BD}}} \times
	\nonumber \\
	&&\hspace{1cm}\times \tan\left(\frac{\sqrt{-9\omega_{BD}-12H_{0}}t}{2\sqrt{\omega_{BD}}}+C\right) \,,
\end{eqnarray}
\begin{equation}
	y_2=\dfrac{\sqrt{3}\,\sqrt{3\,{H_{0}}^{2}\,{\omega_{BD}}^{2}+4\,H_{0}\,\omega_{BD}}-3\,H_{0}\,\omega_{BD}}{\omega_{BD}}\,,
\end{equation}
\begin{equation}
	y_3=-\dfrac{\sqrt{3}\,\sqrt{3\,{H_{0}}^{2}\,{\omega_{BD}}^{2}+4\,H_{0}\,\omega_{BD}}+3\,H_{0}\,\omega_{BD}}{\omega_{BD}}\,,
\end{equation}
where $C$ is an integration constant.  

Integrating these expressions yields three possible forms for the Brans–Dicke scalar field $\phi(t)$:
\begin{equation}
	\phi_1(t) =C_1{e}^{-\frac{2\sqrt{-9{H_{0}}^{2}\omega_{BD}-12H_{0}}\ln\left(\cos\left(\frac{\sqrt{-9\omega_{BD}-12H_{0}}t}{2\,\sqrt{\omega_{BD}}}+c\right)\right)}{\sqrt{-9\omega_{BD}-12H_{0}}}-3H_{0}t}\,,
\end{equation}
\begin{equation}
	\phi_2(t)=C_2\,e^{\dfrac{\sqrt{3}\,\sqrt{3\,{H_{0}}^{2}\,{\omega_{BD}}^{2}+4\,H_{0}\,\omega_{BD}}-3\,H_{0}\,\omega_{BD}}{\omega_{BD}}t}\,,
\end{equation}
\begin{equation}
	\phi_3(t) = C_3\,e^{-\dfrac{\sqrt{3}\,\sqrt{3\,{H_{0}}^{2}\,{\omega_{BD}}^{2}+4\,H_{0}\,\omega_{BD}}+3\,H_{0}\,\omega_{BD}}{\omega_{BD}}t}\,,
\end{equation}
where $c,\, C_1,\, C_2$, and $C_3$ are integration constants.  

These solutions reveal a variety of possible dynamical behaviors for $\phi(t)$ depending on the sign and magnitude of $\omega_{BD}$ and $H_0$. The first solution, $\phi_1(t)$, represents an oscillatory evolution superimposed on an exponential background, while $\phi_2(t)$ and $\phi_3(t)$ correspond to purely exponential growth or decay modes. In the cosmological context, such behaviors can be associated with effective variations of the gravitational “constant” $G_{\text{eff}} \sim 1/\phi(t)$, which can lead to observable consequences during early-universe epochs.

Finally, it is worth noting that an analytical solution to the full system of equations, including Eq.~\eqref{eq:00 component}, is generally unattainable since the scalar fields $\phi$ and $\varphi$ evolve according to distinct dynamical laws and functional dependencies. A simplifying assumption is to consider asymptotic regimes in which one field dominates over the other. In particular, the limit $\phi \gg \varphi$ effectively reduces the dynamics to the standard Brans–Dicke case, while the inverse limit $\varphi \gg \phi$ corresponds to a purely Palatini-dominated regime. These two limiting cases are of particular interest for cosmological applications, as they may provide insights into the roles of each sector during distinct epochs such as inflation or reheating. A detailed analysis of these regimes, however, will be reserved for future work.

\subsection{Matter cases}

The feasibility of obtaining analytical solutions in the presence of matter components is, once again, closely dependent on the specific functional forms chosen for the Hubble parameter $H(t)$, the coupling function $f(\phi)$, and the scalar-field potentials $V(\phi)$ and $U(\varphi)$. In cosmological models that incorporate matter or radiation \cite{Mimoso:1994wn}, the expansion dynamics are often well approximated by a power-law behavior of the scale factor, $a(t)\sim t^{\alpha}$, which naturally emerges in GR for perfect fluids with a barotropic equation of state. In this case, the Hubble parameter is expressed as
\begin{equation}
	H = \alpha t^{-1}\,,
\end{equation}
where the parameter $\alpha$ encodes the dominant matter component -- for instance, $\alpha = 1/2$ corresponds to the radiation-dominated epoch, while $\alpha = 2/3$ represents the matter-dominated era.

Following the same procedure adopted in the de Sitter analysis, one can identify a class of $f(\hat{R})$ functions that admit analytical scalar-field solutions consistent with this background evolution. For the power-law expansion, the corresponding dynamical equation for the Palatini curvature potential takes the form
\begin{equation}\label{eq:matter potential sol}
	6(2\alpha^2-\alpha)\left(g^{-1}\!\left(\frac{df}{d\hat{R}}\right)\right)^{-2} + \mathbf{G}\!\left(g^{-1}\!\left(\frac{df}{d\hat{R}}\right)\right) = \hat{R}\,,
\end{equation}
which is structurally more intricate than its de Sitter counterpart due to the explicit dependence on the power-law index $\alpha$. The term proportional to $(2\alpha^2-\alpha)$ encodes the deviation from exponential expansion, and thus captures the dynamical influence of matter or radiation.

A general analytical solution to Eq.~\eqref{eq:matter potential sol} can be obtained under the assumption that there exists a function
\begin{equation}
	\mathbf{F}\!\left(\frac{df}{d\hat{R}}\right) \equiv 6(2\alpha^2-\alpha)\left(g^{-1}\!\left(\frac{df}{d\hat{R}}\right)\right)^{-2} + \mathbf{G}\!\left(g^{-1}\!\left(\frac{df}{d\hat{R}}\right)\right),
\end{equation}
which is invertible, i.e., for which an inverse function $\mathbf{F}^{-1}$ exists. Under this condition, the general solution for $f(\hat{R})$ can be formally expressed as
\begin{equation}
	f(\hat{R}) = \int \mathbf{F}^{-1}\!\left(\hat{R}\right)\mathrm{d}\hat{R} + C\,,
\end{equation}
where $C$ is an integration constant.

This expression encapsulates the family of $f(\hat{R})$ models that can support analytical scalar-field solutions compatible with a power-law cosmological background. The dependence on $\alpha$ explicitly links the functional form of $f(\hat{R})$ to the dominant matter content of the Universe, thereby providing a unified framework to explore both radiation and matter-dominated eras within the HMPST formalism.\\


A particularly noteworthy and physically relevant class of $f(\hat{R})$ theories that allows for analytical treatment is the power-law class, characterized by the functional dependence
\begin{equation}
	f(\hat{R}) = \left(\frac{\hat{R}}{R_0}\right)^n,
\end{equation}
where $R_0$ is a constant with dimensions of curvature, and $n$ is a dimensionless parameter controlling the deviation from the standard Einstein-Hilbert term.

Substituting this functional form into the scalar-tensor representation of the theory, and recalling that the Palatini scalar field is defined by $\varphi \equiv \mathrm{d}f/\mathrm{d}\hat{R}$, we obtain
\begin{equation}
	\varphi = n R_0^{-n} \hat{R}^{n-1}.
\end{equation}
Inverting this expression gives the Palatini curvature scalar as a function of the scalar field,
\begin{equation}
	\hat{R} = R_0 \left(\frac{\varphi}{n}\right)^{\frac{1}{n-1}}.
\end{equation}
Replacing this into the definition of the potential $U(\varphi) \equiv \hat{R}\varphi - f(\hat{R})$ yields a power-law potential of the form
\begin{equation}
	U(\varphi) = U_0\, \varphi^m,
\end{equation}
where the constants $U_0$ and $m$ are determined by
\begin{equation}
	U_0 \equiv \frac{n - 1}{n^m} R_0^m, 
	\qquad 
	m \equiv \frac{n}{n - 1}.
\end{equation}

This class of models is of particular interest because the parameter $n$ controls the effective dynamics of the scalar field, influencing both the background expansion and the possible cosmological solutions. For instance, positive values of $n>1$ often enhance curvature effects at high energies, leading to early-time inflationary-like behavior, while $n<1$ may induce corrections relevant to late-time cosmic acceleration. Thus, the power-law class provides a flexible yet analytically tractable framework to investigate deviations from GR within the hybrid metric-Palatini approach.

Furthermore, the cosmological equations, namely Eqs.~\eqref{eq:00 component} and~\eqref{eq:phi FLRW}, also admit power-law solutions when the coupling and potential functions appearing in the action take the monomial forms
\begin{equation}
	f(\phi) = f_0 \phi^{n_f}, \quad 
	V(\phi) = V_0 \phi^{n_V}, \quad 
	\omega(\phi) = \omega_0 \phi^{n_\omega},
\end{equation}
where $f_0$, $V_0$, and $\omega_0$ are constant coefficients, and $n_f$, $n_V$, and $n_\omega$ are real exponents characterizing each function. These power-law parametrizations are widely employed in scalar-tensor and modified gravity models since they allow analytical treatments and can describe different cosmological epochs through suitable choices of the exponents.

Starting with Eq.~\eqref{eq:varphi FRLW}, and adopting the power-law scale factor $a(t) \sim t^{\alpha}$, the field equation for the scalar $\varphi$ becomes
\begin{equation}
	6(2\alpha^2 - \alpha)t^{-2} 
	- \frac{3}{2} \left( \frac{\dot{\varphi}}{\varphi} \right)^2 
	+ 9\alpha t^{-1} \frac{\dot{\varphi}}{\varphi} 
	+ 3 \frac{\ddot{\varphi}}{\varphi} 
	- U_0 m \varphi^{m-1} = 0,
\end{equation}
which admits a power-law solution of the form
\begin{equation}
	\varphi(t) = \varphi_0 \, t^{\frac{2}{1 - m}},
\end{equation}
where the constant amplitude $\varphi_0$ is determined by
\begin{equation}
	\varphi_0 = 
	\left[
	\frac{
		6(2\alpha^2 - \alpha) 
		+ \frac{6m}{(1 - m)^2} 
		+ \frac{18\alpha}{1 - m}
	}{U_0 m}
	\right]^{\!\frac{1}{m - 1}}.
\end{equation}

Analogously, for the scalar field $\phi$, Eq.~\eqref{eq:phi FLRW} becomes
\begin{align}
	&6 n_f f_0 (2\alpha^2 - \alpha) \phi^{n_f - 1} t^{-2}
	- \omega_0 (n_\omega - 1) \phi^{n_\omega - 2} \dot{\phi}^2
		\nonumber \\
		&
	- 2\omega_0 \phi^{n_\omega - 1} \ddot{\phi}  - 6\alpha \omega_0 \phi^{n_\omega - 1} t^{-1} \dot{\phi}
	+ V_0 n_V \phi^{n_V - 1} = 0.
\end{align}
Under the condition $n_f = n_\omega + 1$, this equation also admits a power-law solution of the form
\begin{equation}
	\phi(t) = \phi_0 \, t^{\beta},
\end{equation}
where the exponent $\beta$ is given by
\begin{equation}
	\beta = \frac{2}{n_\omega + 1 - n_V},
\end{equation}
and the amplitude $\phi_0$ reads
\begin{equation}
	\phi_0 = 
	\left(
	\frac{V_0 n_V}{
		\bar{\beta}(\beta, \omega_0, \alpha)
	}
	\right)^{\!-\frac{1}{n_V - n_\omega - 1}},
\end{equation}
with the auxiliary function $\bar{\beta}$ introduced for notational simplicity as
\begin{eqnarray}
	\bar{\beta} & \equiv & 
	6 f_0 (-n_\omega - 1)(2\alpha^2 - \alpha)
	- \omega_0 (1 - n_\omega) \beta^2
		\nonumber \\
	&& + 2\omega_0 \beta (\beta - 1)
	+ 6 \omega_0 \alpha \beta.
\end{eqnarray}

Applying now these results in the Friedmann equation~\eqref{eq:00 component}, we obtain
\begin{eqnarray}
	&& 3\alpha^2 \left[ f_0^{n_\omega+1} \phi_0^{n_\omega+1} 
	t^{\frac{2n_\omega+2}{n_\omega+1-n_V}-2} 
	+ \lambda \varphi_0 t^{\frac{2}{1-m}-2} \right] \nonumber \\
	&& + 3\alpha \left[ f_0^{n_\omega+1} 
	\frac{2n_\omega+2}{n_\omega+1-n_V} 
	t^{\frac{2n_\omega+2}{n_\omega+1-n_V} - 2} 
	+ \lambda \varphi_0 \frac{2}{1-m} 
	t^{\frac{2}{1-m} - 2} \right] \nonumber \\
	&& + \frac{3}{4} \lambda 
	\left( \frac{2}{1-m} \right)^2 
	t^{\frac{2}{1-m}-2} 
	- \frac{{\omega_0} \phi_0^{n_\omega+1} 
		t^{\frac{2n_\omega+2}{n_\omega+1-n_V}}}{n_\omega+1-n_V} \nonumber \\
	&& + \frac{1}{2} V_0 \phi_0 
	t^{\frac{2n_V}{n_\omega+1-n_V}} 
	- \frac{\lambda}{2} U_0 \varphi_0 
	t^{\frac{2m}{1-m}} 
	= \kappa^2 \rho\,,
\end{eqnarray}
where $\rho$ denotes the energy density of the matter content. For the particular case in which the potential exponents satisfy the relation
\begin{equation}
	n_V = (n_\omega + 1) m,
\end{equation}
and by assuming that the matter density evolves according to the continuity equation,
\begin{equation}
	\rho = \rho_0 \, t^{-3\alpha(1+w)}\,,
\end{equation}
corresponding to the general solution of Eq.~\eqref{eq:continuity} for a barotropic fluid with constant equation-of-state parameter $w$, one can equate the powers of $t$ on both sides of the Friedmann equation. This procedure yields the consistency condition
\begin{equation}\label{eq:alpha power law}
	\alpha = \frac{2m}{3(m - 1)(1 + w)} = n \, \alpha_{\text{GR}}\,,
\end{equation}
where $\alpha_{\text{GR}} = \frac{2}{3(1+w)}$ represents the standard GR result for a perfect-fluid dominated Universe, and $n$ acts as an effective scaling factor that encapsulates the deviation from GR introduced by the hybrid metric–Palatini framework.

Substituting this result back into the Friedmann equation determines the matter density normalization as
\begin{eqnarray}
	\kappa^2\rho_0 &=& 3\alpha^2 \left[ f_0^{n_\omega+1} \phi_0^{n_\omega+1}  
	+ \lambda \varphi_0 \right] \nonumber \\
	&& + 3\alpha \left[ f_0^{n_\omega+1} 
	\frac{2n_\omega+2}{n_\omega+1-n_V}
	+ \frac{2\lambda \varphi_0}{1 - m} \right] \nonumber \\
	&& + \frac{3}{4} \lambda 
	\left( \frac{2}{1 - m} \right)^2  
	- \frac{\omega_0 \phi_0^{n_\omega+1}}{n_\omega+1 - n_V} \nonumber \\
	&& + \frac{1}{2} V_0 \phi_0  
	- \frac{\lambda}{2} U_0 \varphi_0\,.
\end{eqnarray}

The central outcome of this class of power-law models is encapsulated in Eq.~\eqref{eq:alpha power law}, which quantifies the deviation from the standard cosmological dynamics predicted by GR. Specifically, the modification introduced by the Palatini sector directly affects the evolution of the scale factor, and this deviation is entirely controlled by the parameter $n$, which arises from the power-law form of $f(\hat{R})$. In this sense, the exponent $n$ serves as an effective measure of how strongly the model departs from standard GR behavior, acting as a key indicator of the influence of the Palatini contribution on the overall cosmological dynamics.

It is important to emphasize, however, that Eq.~\eqref{eq:alpha power law} and the associated cosmological solutions derived for this class of models are not valid for the special case $n = 1$. In this limit, the Palatini sector becomes linear in $\hat{R}$, reducing the theory to a scalar-tensor framework with an effective cosmological constant, as described in Eq.~\eqref{eq:linear palatini action}. This scenario is characterized by a different set of field equations and cosmological behavior, and therefore cannot be recovered by naively setting $n = 1$ in the power-law expressions.

Nevertheless, models with values of $n$ close to unity, \textit{i.e.}, $n \approx 1$, provide a controlled departure from GR, allowing one to systematically explore modifications to standard cosmology while maintaining an evolution of the scale factor that remains close to the familiar behavior predicted by GR. Such small deviations offer a physically motivated framework to investigate the impact of the Palatini sector on early- and late-time cosmological dynamics, including possible contributions to accelerated expansion or inflationary behavior.

\section{Vacuum Solutions}\label{SectionVII}

We now turn our attention to spherically symmetric and static configurations within the context of vacuum solutions to the generalized Palatini field equations. In full generality, the structure of these solutions is highly sensitive to the specific functional forms of $f(\hat{R})$ and $f(\phi)$, as well as the scalar potential $V(\phi)$. Obtaining explicit analytic solutions for arbitrary choices of these functions is, in general, a formidable task due to the nonlinear and coupled nature of the resulting field equations. Consequently, to gain qualitative and quantitative insight into the underlying dynamics of the theory, it is both natural and productive to focus on a tractable subclass of models. Such a subclass should ideally retain the essential physical characteristics of the general theory while enabling meaningful analytical development.

A convenient and physically meaningful simplification is to consider a linear form of the function $f(\hat{R})$. This choice reduces the complexity of the field equations significantly, while still capturing nontrivial interactions between the scalar and metric sectors. By imposing this restriction, we are able to recast the theory in a form amenable to analytical techniques, and, crucially, to identify exact solutions that can serve as benchmarks for more general explorations.

Concretely, we start from the action in Eq.~\eqref{eq:linear palatini action} with vanishing matter contributions, $S_m = 0$, and perform a conformal transformation to the Einstein frame, as outlined in Sec.~\ref{SectionV}. The resulting action takes the form
\begin{equation}\label{eq:Einstein-scalar gravity}
	S = \frac{1}{2\kappa^2} \int \mathrm{d}^4x \sqrt{-\tilde{g}} \left[ \tilde{R} - \tilde{\omega}(\phi) \tilde{g}^{\mu\nu} \partial_\mu \phi \partial_\nu \phi - \tilde{V}(\phi) \right],
\end{equation}
where the redefined kinetic coupling and potential are given by
\begin{align}
	\tilde{\omega}(\phi) &= \frac{3}{2} \left( \frac{f'(\phi)}{f(\phi)+\lambda \varphi_0} \right)^2 + \frac{\omega(\phi)}{\phi \,[f(\phi)+\lambda \varphi_0]}, \\
	\tilde{V}(\phi) &= \frac{V(\phi) + U_0}{\left( f(\phi) + \lambda \varphi_0 \right)^2}.
\end{align}
Here, the conformal transformation effectively decouples the scalar curvature from the scalar field dynamics, recasting the theory as a scalar–tensor system in the Einstein frame. This representation is particularly convenient because it allows the identification of solution classes familiar from standard Einstein-scalar gravity.

A notable and analytically tractable scenario arises when the effective kinetic coupling $\tilde{\omega}(\phi)$ is constant, $\tilde{\omega}(\phi) = \tilde{\omega}_0$, and the potential is either zero, $\tilde{V} = 0$, or a cosmological constant, $\tilde{V} = \Lambda$. In the case $\tilde{V} = 0$, the system reduces precisely to Einstein-scalar gravity with a minimally coupled massless scalar field. This theory admits the well-known Janis--Newman--Winicour (JNW) solutions~\cite{Janis:1968zz}, which describe static, spherically symmetric spacetimes sourced by a scalar field. Historically, the earliest derivation of these solutions was provided by Fisher in 1948~\cite{Fisher:1948yn}, and subsequent works by Buchdahl~\cite{Buchdahl:1959nk} and Wyman~\cite{Virbhadra:1997ie} have confirmed their equivalence to the JNW configuration~\cite{Bhadra:2001fx}. 

Therefore, the linear-$f(\hat{R})$ subclass of the HMPST theory admits these classical solutions, providing a concrete arena in which the interplay between scalar fields and geometry can be explored explicitly. This identification not only serves as a consistency check for the generalized framework but also highlights the physical relevance of the subclass, as it captures essential scalar-metric interactions while remaining analytically solvable. In this sense, the linear-$f(\hat{R})$ models act as a bridge between fully general, intractable Palatini theories and the well-understood landscape of Einstein-scalar gravity solutions.

When the effective potential $\tilde{V}$ is taken to be constant, $\tilde{V} = \Lambda$, the resulting configurations generalize the classical Buchdahl solution. This scenario was systematically analyzed in~\cite{Pereira:2024olv}, where it was shown that the metric $\tilde{g}_{\mu\nu}$, which satisfies the field equations derived from Eq.~\eqref{eq:Einstein-scalar gravity}, can be expressed in the form
\begin{equation}
	\tilde{g}_{\mu\nu} = \bigl((-\bar{g}_{00})^{1-\beta}, (-\bar{g}_{00})^{\beta} \bar{g}_{ij} \bigr)\,,
\end{equation}
where
\begin{equation}
	\beta = 1 \pm \sqrt{1 + 2 \tilde{\omega}_0 k^2}\,,
\end{equation}
and $k$ is an arbitrary constant characterizing the scalar field amplitude. Here, $\bar{g}_{\mu\nu}$ denotes any cyclic metric with signature $(-,+,+,+)$ that solves the Einstein equations with a cosmological constant:
\begin{equation}
	R_{\mu\nu} = \Lambda \bar{g}_{\mu\nu}.
    \label{labelLambda}
\end{equation}
This condition corresponds to the standard maximally symmetric vacuum equations of GR. 

The scalar field sector associated with these solutions admits the simple logarithmic form
\begin{equation}
	\phi = k \ln(-\bar{g}_{00})\,,
\end{equation}
which ensures consistency with the constant kinetic coupling condition $\tilde{\omega} = \tilde{\omega}_0$ and allows the full set of Einstein-scalar equations to be satisfied.

To illustrate this construction, consider the well-known Schwarzschild--de Sitter (SdS) metric, which is a solution of Eq.~(\ref{labelLambda}) above with $\Lambda \neq 0$:
\begin{align}
	\mathrm{d}s^2 &= -\left(1 - \frac{2GM}{r} - \frac{\Lambda r^2}{3}\right) \mathrm{d}t^2 \nonumber \\
	&\quad + \frac{\mathrm{d}r^2}{1 - \frac{2GM}{r} - \frac{\Lambda r^2}{3}} + r^2 \mathrm{d}\Omega^2,
\end{align}
where $\mathrm{d}\Omega^2$ is the standard angular line element.  

In the Einstein frame, the corresponding solution for the HMPST theory takes the form
\begin{align}
	&\mathrm{d}s^2_{\text{EF}} = -\left(1 - \frac{2GM}{r} - \frac{\Lambda r^2}{3} \right)^{1-\beta} \mathrm{d}t^2 
		\nonumber  \\
	& \quad + \left(1 - \frac{2GM}{r} - \frac{\Lambda r^2}{3} \right)^{\beta} \Biggl[ \frac{\mathrm{d}r^2}{1 - \frac{2GM}{r} - \frac{\Lambda r^2}{3}} + r^2 \mathrm{d}\Omega^2 \Biggr], 
\end{align}
with the corresponding Jordan-frame metric obtained via the conformal transformation
\begin{equation}
	g_{\mu\nu}^{\text{JF}} = \bigl(f(\phi) + \lambda \varphi_0\bigr) g_{\mu\nu}^{\text{EF}},
\end{equation}
and the scalar field profile given by
\begin{equation}
	\phi(r) = k \ln \left[\left(1 - \frac{2GM}{r} - \frac{\Lambda r^2}{3} \right)^{1-\beta} \right].
\end{equation}

An important subtlety is that in the original Jordan frame, both $f(\phi)$ and $\omega(\phi)$ are a priori undetermined functions. Their forms are constrained dynamically by the requirement that the Einstein-frame kinetic coupling $\tilde{\omega}$ remains constant:
\begin{equation}
	\tilde{\omega}(\phi) = \tilde{\omega}_0.
\end{equation}
This condition leads to the differential constraint
\begin{equation}
	3 (f')^2 \phi + 2 (f(\phi) + \lambda \varphi_0) \Bigl[\omega(\phi) - \phi \tilde{\omega}_0 (f(\phi) + \lambda \varphi_0) \Bigr] = 0.
\end{equation}

This equation can be solved in two complementary ways. If a functional form for $\omega(\phi)$ is chosen, then $f(\phi)$ is determined by integrating the differential equation. Conversely, if $f(\phi)$ is specified, then the corresponding form of $\omega(\phi)$ is
\begin{equation}
	\omega(\phi) = \phi \bigl(f(\phi) + \lambda \varphi_0\bigr) \left[ \tilde{\omega}_0 - \frac{3}{2} \left( \frac{f'(\phi)}{f(\phi) + \lambda \varphi_0} \right)^2 \right].
\end{equation}
This explicit relation guarantees that the Einstein-frame kinetic coupling is constant and ensures consistency between the Jordan- and Einstein-frame descriptions of the theory.

\section{Weak-field, slow-motion behavior}\label{SectionVIII}

The influence of scalar fields on Solar--System dynamics can be systematically analyzed in the weak-field, post-Newtonian regime. In this limit, the gravitational field is only slightly perturbed from a fixed background, and deviations from flat spacetime (or from a maximally symmetric background) can be treated perturbatively. To implement this, we linearize the theory by expanding all fields around a chosen background solution 
\((\bar g_{\mu\nu},\phi_0,\varphi_0)\) of the full field equations. Here, $\bar g_{\mu\nu}$ represents the background metric, while $\phi_0$ and $\varphi_0$ denote the constant background values of the scalar fields.  

Indices are raised and lowered with the background metric $\bar g_{\mu\nu}$, and all covariant derivatives are defined with respect to its Levi–Civita connection $\bar\nabla_\mu$. To remain within the linearized regime, we retain only terms that are first order in the perturbations, neglecting quadratic and higher-order contributions such as
\begin{equation}
	\mathcal{O}\bigl(h^2,\,\delta\phi^2,\,\delta\varphi^2,\,h\,\delta\phi,\,h\,\delta\varphi,\,\delta\phi\,\delta\varphi\bigr),
\end{equation}
where $h_{\mu\nu}$, $\delta\phi$, and $\delta\varphi$ represent small deviations from the background.

Accordingly, the metric and scalar fields are decomposed as
\begin{align}
	g_{\mu\nu} &= \bar g_{\mu\nu} + h_{\mu\nu}, \qquad |h_{\mu\nu}| \ll 1,\\
	\phi &= \phi_0 + \delta\phi, \qquad |\delta\phi| \ll 1,\\
	\varphi &= \varphi_0 + \delta\varphi, \qquad |\delta\varphi| \ll 1.
\end{align}

Of particular interest is the composite function
\begin{equation}
	F(\phi,\varphi) \equiv f(\phi) + \lambda \varphi,
\end{equation}
which appears frequently in the field equations. Expanding $F(\phi,\varphi)$ to first order about the background yields
\begin{equation}
	F = F_0 + \delta F,
\end{equation}
where
\begin{equation}
	F_0 \equiv f(\phi_0) + \lambda\,\varphi_0, \qquad 
	\delta F \equiv f_\phi(\phi_0)\,\delta\phi + \lambda\,\delta\varphi.
\end{equation}

This linearized expansion forms the basis for computing post-Newtonian corrections to gravitational observables. In particular, it allows one to express the modifications to the effective gravitational coupling and the scalar-mediated interactions in terms of the small perturbations $(h_{\mu\nu}, \delta\phi, \delta\varphi)$, facilitating direct comparison with Solar–System experiments.

For Solar--System applications, it is natural to choose an asymptotically flat background,
\begin{equation}
	\bar g_{\mu\nu} = \eta_{\mu\nu},
\end{equation}
and to consider the static regime, in which time derivatives can be neglected, i.e., $\Box \longrightarrow \vec{\nabla}^2$.

Starting from the background equations, namely Eqs.~\eqref{eq:FE special case phi}, \eqref{eq:FE special case varphi} and \eqref{eq:trace metric}, one finds that the background values of the potentials satisfy
\begin{equation}
	V(\phi_0) = \lambda U(\varphi_0), \qquad 
	V_\phi(\phi_0) = U_\varphi(\varphi_0) = 0.
\end{equation}
These conditions ensure that the chosen background is indeed a solution of the field equations.

To linear order, the trace of the metric equation and the scalar-field equations reduce to the coupled system
\begin{align}
	\delta R &= \frac{3}{F_0} \vec{\nabla}^2 \delta F + \frac{\kappa^2}{F_0} \rho, \\
	\bigl(\vec{\nabla}^2 - m_\phi^2 \bigr) \delta \phi &= -\alpha\, \delta R, \\
	\bigl(\vec{\nabla}^2 - m_\varphi^2 \bigr) \delta \varphi &= -\beta\, \delta R,
\end{align}
where the constants are defined as
\begin{eqnarray}
	\alpha \equiv \frac{\phi_0 f_\phi(\phi_0)}{2 \omega(\phi_0)}, \qquad
	&& \beta \equiv \frac{\varphi_0}{3}, \qquad
	m_\phi^2 \equiv -\phi_0 \frac{V_{\phi\phi}(\phi_0)}{2 \omega(\phi_0)}, 
		\nonumber \\
	&& m_\varphi^2 \equiv -\frac{\varphi_0 U_{\varphi\varphi}(\varphi_0)}{3}.
\end{eqnarray}

These are Helmholtz-type equations with a common source term $\delta R$. For a point mass source, $\rho = M \, \delta^3(\vec{x})$, the general solutions for the scalar-field perturbations take the form
\begin{align}
	\delta \phi(r) &= \frac{\kappa^2 M}{4\pi F_0} \frac{\alpha}{A_2} \frac{1}{r} 
	\Biggl[ 
	\frac{\mu_+^2 - m_\varphi^2}{\mu_+^2 - \mu_-^2} e^{-\mu_+ r} 
	+ \frac{m_\varphi^2 - \mu_-^2}{\mu_+^2 - \mu_-^2} e^{-\mu_- r} 
	\Biggr], \\
	\delta \varphi(r) &= \frac{\kappa^2 M}{4\pi F_0} \frac{\beta}{A_2} \frac{1}{r} 
	\Biggl[ 
	\frac{\mu_+^2 - m_\phi^2}{\mu_+^2 - \mu_-^2} e^{-\mu_+ r} 
	+ \frac{m_\phi^2 - \mu_-^2}{\mu_+^2 - \mu_-^2} e^{-\mu_- r} 
	\Biggr],
\end{align}
where
\begin{equation}
	\mu_\pm^2 = \frac{A_1 \pm \sqrt{A_1^2 - 4 A_2 A_0}}{2 A_2},
\end{equation}
and
\begin{align}
	A_2 &= 1 + \frac{3}{F_0} \left( f_\phi(\phi_0)\, \alpha + \lambda \, \beta \right), \\
	A_1 &= m_\phi^2 + m_\varphi^2 + \frac{3}{F_0} \left( f_\phi(\phi_0) \, \alpha \, m_\varphi^2 + \lambda \, \beta \, m_\phi^2 \right), \\
	A_0 &= m_\phi^2 \, m_\varphi^2.
\end{align}

From these scalar-field solutions, the perturbation of the composite function $F$ is
\begin{equation}
	\delta F(r) = \frac{\kappa^2 M}{4\pi F_0} \frac{1}{r A_2} \left( A_+ e^{-\mu_+ r} - A_- e^{-\mu_- r} \right),
\end{equation}
with
\begin{equation}
	A_\pm = \frac{ f_\phi(\phi_0) \, \alpha \, (\mu_\pm^2 - m_\varphi^2) + \lambda \, \beta \, (\mu_\pm^2 - m_\phi^2)}{\mu_+^2 - \mu_-^2}.
\end{equation}

These expressions explicitly demonstrate that, in the weak-field, static limit, the scalar fields and the effective gravitational coupling $\delta F$ acquire Yukawa-type corrections, with characteristic length scales set by the masses $\mu_\pm^{-1}$. This forms the basis for computing post-Newtonian corrections and comparing the theory's predictions with Solar--System tests.

Using now Eq.~\eqref{eq:FE special case metric} and its trace, the linearized metric equation can be written as
\begin{equation}\label{eq:perturbed Ricci}
	F_0 \, \delta R_{\mu\nu} = \bar{\nabla}_\mu \bar{\nabla}_\nu \delta F + \frac{1}{2} \bar{g}_{\mu\nu} \, \bar{\Box} \delta F + \kappa^2 \left( T_{\mu\nu} - \frac{1}{2} \bar{g}_{\mu\nu} T \right),
\end{equation}
where $\delta R_{\mu\nu}$ is the perturbation of the Ricci tensor and $T_{\mu\nu}$ is the stress-energy tensor of matter.

For a Minkowski background, the linearized Ricci tensor takes the standard form
\begin{equation}
	\delta R_{\mu\nu} = \frac{1}{2} \left( \partial_\mu \partial_\sigma \tilde{h}_\nu^\sigma + \partial_\nu \partial_\sigma \tilde{h}_\mu^\sigma - \Box h_{\mu\nu} \right),
\end{equation}
where 
\begin{equation}
	\tilde{h}_\nu^\sigma \equiv h_\nu^\sigma - \frac{1}{2} \delta_\nu^\sigma h,
\end{equation}
and $h \equiv \eta^{\mu\nu} h_{\mu\nu}$ is the trace of the metric perturbation.

To simplify the linearized equations, we impose the generalized harmonic gauge
\begin{equation}
	\partial_\sigma \tilde{h}^{\sigma}_\mu - \frac{\partial_\mu \delta F}{F_0} = 0,
\end{equation}
which generalizes the standard harmonic gauge to accommodate the scalar-field contribution $\delta F$.

Substituting this gauge condition into Eq.~\eqref{eq:perturbed Ricci} and considering the static regime ($\Box \rightarrow \vec{\nabla}^2$) yields
\begin{equation}
	-\frac{1}{2} \vec{\nabla}^2 h_{\mu\nu} = \frac{\bar{g}_{\mu\nu}}{2 F_0} \vec{\nabla}^2 \delta F + \frac{\kappa^2}{F_0} \left( T_{\mu\nu} - \frac{1}{2} \bar{g}_{\mu\nu} T \right).
\end{equation}

For a point-mass source, $\rho = M \delta^3(\vec{x})$, the solution of this equation gives the post-Newtonian metric components
\begin{align}
	h_{00}(r) &= \frac{2 G_{\text{eff}} M}{r}, \\
	h_{ij}(r) &= \frac{2 \gamma \, G_{\text{eff}} M}{r} \, \delta_{ij},
\end{align}
where the effective gravitational constant and the post-Newtonian parameter $\gamma$ are
\begin{align}
	G_{\text{eff}} &= \frac{G}{F_0} \left[ 1 + \epsilon(r) \right], \\
	\gamma &= \frac{1 - \epsilon(r)}{1 + \epsilon(r)},
\end{align}
with
\begin{equation}
	\epsilon(r) \equiv \frac{1}{A_2 F_0} \left( A_+ e^{-\mu_+ r} - A_- e^{-\mu_- r} \right),
\end{equation}
where $A_\pm$ and $\mu_\pm$ are defined above.

These expressions clearly show that the deviation from GR depends on the background values of the scalar-field parameters. In particular, when the scalar-field masses are large ($m_\phi, m_\varphi \gg 1$), implying $\mu_\pm \gg 1$, and for $F_0 \approx 1$, one recovers
\begin{equation}
	G_{\text{eff}} \approx G, \qquad \gamma \approx 1,
\end{equation}
so the theory effectively reduces to GR on Solar-System scales.

Even for light scalar fields, the deviation can remain small if the linear couplings in $F$ are sufficiently weak:
\begin{equation}
	\Big| \frac{\phi_0 f_\phi(\phi_0)^2}{2 \, \omega(\phi_0) \, F_0} \Big| \ll 1, 
	\qquad
	\Big| \frac{\lambda \, \varphi_0}{3 \, F_0} \Big| \ll 1.
	\label{eq:weak-mix-bounds}
\end{equation}
Under these conditions, $A_2 \simeq 1$ and $|A_\pm| \ll \mu_\pm^2$, so that $|\epsilon(r)| \ll 1$ for all distances $r$, ensuring that Solar-System constraints are satisfied.  

This analysis demonstrates that the HMPST theory can naturally satisfy current observational bounds while allowing for nontrivial scalar-field dynamics at larger scales.

\section{Discussion of the approach}\label{Section:Discussion}

While GR is remarkably successful in all regimes where it has been experimentally tested, this success does not imply that gravity is fundamentally exhausted by the Einstein-Hilbert action at all curvature or energy scales. From an effective field-theory perspective, GR should be regarded as the leading low-energy description of a more general gravitational dynamics, with additional degrees of freedom becoming relevant only in specific curvature or density windows. The HMPST framework explored in this work is constructed precisely in this spirit. The Einstein-Hilbert term ensures that the theory reproduces the standard strong-field and local behavior of GR, while the Palatini sector introduces an additional propagating scalar degree of freedom that can modify gravity in a controlled and observationally viable manner.
	
A key result of the present paper is that the linear-$f(\hat R)$ HMPST subclass admits a clear separation between regimes where modifications are active and regimes where GR is recovered. In the weak-field limit, the linearized theory around Minkowski spacetime leads to Yukawa-type corrections to the Newtonian potential, characterized by interaction ranges $\mu_\pm^{-1}$ and an effective Newton constant $G_{\rm eff}$. As explicitly shown in Sec.~VIII, Solar-System consistency is automatically achieved when the scalar modes are sufficiently massive or weakly coupled, yielding $G_{\rm eff}\!\to\!G$ and the post-Newtonian parameter $\gamma\!\to\!1$. Thus, the theory does not compete with GR in the regimes where GR is known to work extremely well, but instead extends it in a technically natural way that preserves all current local constraints.
		
The presence of two scalar fields is not an arbitrary complication but a physically well-motivated feature. The Palatini-induced scalar $\varphi$ arises unavoidably from the scalar-tensor representation of the hybrid metric-Palatini sector and controls departures from GR through its potential $U(\varphi)$. Its dynamics are directly responsible for the modified gravitational interaction and its screening in high-density environments. The second scalar field $\phi$ is introduced as a modulator of the curvature couplings, a structure that is common in scalar-tensor effective actions and in broader high-energy frameworks where multiple scalar degrees of freedom naturally emerge, such as dilaton or moduli sectors in compactified theories~\cite{Grana:2005jc}. Within the HMPST framework, $\phi$ provides an additional handle on the cosmological dynamics without spoiling the GR limit.

From a phenomenological standpoint, the two scalars play complementary roles. Through the functions $f(\phi)$ and $V(\phi)$, the field $\phi$ can act as an inflaton-like degree of freedom in the early Universe or as a quintessence-like field driving late-time acceleration, as explicitly illustrated by the existence of de Sitter and matter-dominated solutions in the homogeneous and isotropic sector. At the same time, the scalar $\varphi$ governs the strength and range of deviations from GR in both cosmological and astrophysical settings. This naturally places the theory within the broader class of multi-field scalar-tensor models, where well-known mechanisms such as hybrid-inflation-type evolution~\cite{Linde:1993cn,Linde:1991km}, energy transfer and reheating~\cite{Kofman:1994rk,Kofman:1997yn}, and the presence of adiabatic and entropy modes~\cite{Gordon:2000hv,Wands:2007bd} can be consistently realized, depending on the choice of potentials and couplings.
	
Equally important is the question of energy scales. The HMPST theory is not intended as a modification of gravity operative at all times and scales. Rather, it should be interpreted as an effective description whose domain of applicability is set by the masses and couplings of the scalar sector. In practice, this means that deviations from GR can be entirely negligible throughout most of the cosmic history and in high-density environments, while becoming dynamically relevant only when curvature or energy densities fall within the window where the scalar fields are sufficiently light to evolve cosmologically. This scale-dependent behavior is explicitly encoded in the analytic weak-field solutions and in the cosmological sector discussed in this work.
	
In this sense, the HMPST framework does not aim to replace GR, but to extend it in a theoretically controlled and phenomenologically flexible manner. By admitting exact cosmological and spherically symmetric solutions, recovering GR in the appropriate limits, and providing a clear interpretation of its additional degrees of freedom, the linear-$f(\hat R)$ HMPST subclass offers a coherent arena for exploring gravitational physics beyond GR while remaining firmly anchored to observational viability.

\section{Summary and Conclusions}\label{Section:Conclusion}

Motivated by the success of hybrid metric–Palatini gravity in unifying metric and Palatini contributions while preserving a dynamical scalar degree of freedom, it is natural to explore further generalizations in which an explicit scalar field $\phi$ modulates the curvature couplings. This construction enhances the theoretical flexibility of the model, enabling richer cosmological dynamics, control over the scalar kinetic sector, and the inclusion of nontrivial self-interactions through appropriate potentials. By encompassing both the metric and Palatini limits as particular cases, this framework provides a variationally consistent and conceptually unified extension of previously studied scalar-tensor theories. 

In this work, we have undertaken a detailed study of a specific, analytically tractable realization of this broader framework: the linear-$f(\hat{R})$ subclass of hybrid metric-Palatini scalar-tensor (HMPST) theories. Our analysis spans cosmological, strong-field, and weak-field regimes, emphasizing the interplay between the metric and scalar sectors governed by the functions $f(\hat{R})$, $f(\phi)$, and the scalar potentials $V(\phi)$ and $U(\varphi)$. Given the complexity of the general field equations, this subclass serves as a consistent and insightful model that retains the essential physical features of scalar-metric couplings while allowing for explicit analytic solutions and a phenomenological interpretation.

We first explored cosmological solutions, considering homogeneous and isotropic spacetimes. In particular, we analyzed de Sitter solutions, corresponding to late-time accelerated expansion, and matter-dominated solutions relevant for the early universe. These configurations illustrate how the scalar fields in HMPST theories can act as effective sources driving cosmic acceleration or influencing structure formation. The flexibility in the choice of scalar potentials and couplings allows these solutions to interpolate between different cosmological epochs, providing a natural connection to dark-energy phenomenology.

Building on this cosmological foundation, we investigated spherically symmetric, static vacuum configurations. By restricting to a linear form of $f(\hat{R})$, the field equations simplify, permitting analytic solutions while retaining nontrivial scalar-metric dynamics. Transforming to the Einstein frame, the theory is expressed as an effective scalar-tensor system with kinetic coupling $\tilde{\omega}(\phi)$ and potential $\tilde{V}(\phi)$. For constant $\tilde{\omega}=\tilde{\omega}_0$ and either vanishing or constant potential, the theory admits well-known analytic solutions: (i) $\tilde{V}=0$ reproduces the Janis-Newman-Winicour (JNW) solutions and their equivalents (Fisher, Buchdahl, Wyman), describing static, massless-scalar spacetimes.  (ii) $\tilde{V}=\Lambda$ constant generalizes the Buchdahl solution, including the Schwarzschild-de Sitter metric in the Einstein frame with logarithmic scalar-field profiles, while the Jordan-frame metric is obtained via the conformal factor. The functions $f(\phi)$ and $\omega(\phi)$ are dynamically constrained to satisfy $\tilde{\omega}=\tilde{\omega}_0$, preserving analytic tractability.

In the weak-field regime, relevant to Solar-System tests, we linearized the metric and scalar fields around an asymptotically flat Minkowski background. The perturbations satisfy coupled Helmholtz-type equations sourced by matter, yielding Yukawa-type corrections to the scalar fields and the effective gravitational coupling $F(\phi,\varphi)$. Employing a generalized harmonic gauge, we derived the post-Newtonian metric, obtaining explicit expressions for the effective gravitational constant $G_{\rm eff}$ and the post-Newtonian parameter $\gamma$. Deviations from GR are controlled by the scalar-field masses, background values, and the linear couplings in $F$. In the limits of heavy scalar fields or small couplings, $G_{\rm eff}\approx G$ and $\gamma\approx 1$, ensuring consistency with Solar-System observations. Even for light scalars, small couplings suppress deviations, guaranteeing phenomenological viability.

Taken together, these results illustrate the versatility of the linear-$f(\hat{R})$ HMPST subclass: it provides analytic cosmological and strong-field solutions while simultaneously accommodating controlled deviations from GR in the weak-field limit. The theory can consistently interpolate between cosmological, astrophysical, and local scales, highlighting its potential to unify multiple gravitational regimes within a single framework.

Looking ahead, several directions are particularly promising. Extending the analysis to dynamical and rotating spacetimes would generalize static solutions to realistic astrophysical objects such as rotating stars or black holes. Cosmological studies of perturbations could constrain scalar-field parameters through observations of the cosmic microwave background, large-scale structure, or late-time acceleration. High-precision astrophysical and gravitational-wave observations offer further opportunities to test the scalar couplings and masses. 

Studies of stability, nonlinear dynamics, and potential signatures in black hole shadows, lensing, or accretion flows would complement these investigations. Finally, the inherent flexibility in the choice of potentials and couplings positions hybrid metric-Palatini theories as a fertile framework for connecting modified gravity with effective field theories, dark-matter phenomenology, or quantum-gravity-inspired extensions. Collectively, these avenues highlight the broad relevance and rich phenomenology of HMPST theories across multiple scales in gravitational physics.

\begin{acknowledgments}
	DSP, FSNL and JPM acknowledge funding from
	the Fundação para a Ciência e a Tecnologia (FCT) through
	the research grants UID/04434/2025 and
	PTDC/FIS-AST/0054/2021. FSNL also acknowledges support from the 
	FCT Scientific Employment Stimulus contract with reference
	CEECINST/00032/2018.
	SC acknowledges the Istituto Nazionale di Fisica Nucleare (INFN) Sez. di Napoli, Iniziative Specifiche QGSKY and MoonLight-2  and the Istituto Nazionale di Alta Matematica (INdAM), gruppo GNFM, for the support. This paper is based upon work from COST Action CA21136 -- Addressing observational tensions in cosmology with systematics and fundamental physics (CosmoVerse), supported by COST (European Cooperation in Science and Technology).	
	
\end{acknowledgments}


\end{document}